\documentclass[preprint2,numberedappendix]{emulateapj}

\newcommand{\OII}{\ensuremath{{\rm [OII] \lambda 3726,3729} }}

\newcommand{\Msun}{\ensuremath{M_{\odot}}}
\newcommand{\Mdyn}{\ensuremath{M_\mathrm{dyn}}}

\newcommand{\Mstar}{\ensuremath{M_{\ast}}}
\newcommand{\HST}{\emph{HST}}
\newcommand{\spitzer}{\emph{Spitzer}}

\newcommand{\R}{\ensuremath{R_e}}
\newcommand{\Sersic}{S{\'e}rsic}
\newcommand{\sigmainf}{\ensuremath{\sigma_\mathrm{inf}}}
\newcommand{\sigmae}{\ensuremath{\sigma_e}}

\usepackage{capt-of}
\usepackage{graphicx}
\usepackage{subfigure}
\usepackage{verbatim}

\shorttitle{Velocity Dispersions for $z>1$ Quiescent Galaxies}
\shortauthors{Belli, Newman and Ellis}
\submitted{Accepted for publication in the Astrophysical Journal}

\begin{document}

\title{Velocity Dispersions and Dynamical Masses for a Large Sample of Quiescent Galaxies at $z>1$: Improved Measures of the Growth in Mass and Size}

\author{Sirio Belli\altaffilmark{1}, Andrew B. Newman\altaffilmark{1,2}, Richard S. Ellis\altaffilmark{1}}
\altaffiltext{1}{Department of Astronomy, California Institute of Technology, MS 249-17, Pasadena, CA 91125, USA}
\altaffiltext{2}{The Observatories of the Carnegie Institution for Science, 813 Santa Barbara St., Pasadena, CA 91101, USA}

\begin{abstract}

We present Keck LRIS spectroscopy for a sample of 103 massive ($M > 10^{10.6} \Msun$) galaxies with redshifts $0.9<z<1.6$. Of these, 56 are quiescent with high signal-to-noise absorption line spectra, enabling us to determine robust stellar velocity dispersions for the largest sample yet available beyond a redshift of 1. Together with effective radii measured from deep \emph{Hubble Space Telescope} images, we calculate dynamical masses and address  key questions relating to the puzzling size growth claimed by many observers for quiescent galaxies over the redshift interval $0<z<2$. Our large sample provides the first opportunity to carefully examine the relationship between stellar and dynamical masses at high redshift. We find this relation closely follows that determined locally. We also confirm the utility of the locally-established empirical calibration which enables high-redshift velocity dispersions to be estimated photometrically, and we determine its accuracy to be 35\%. To address recent suggestions that progenitor bias --- the continued arrival of recently-quenched larger galaxies --- can largely explain the size evolution of quiescent galaxies, we examine the growth at fixed velocity dispersion assuming this quantity is largely unaffected by the merger history. Using the velocity dispersion - age relation observed in the local universe, we demonstrate that significant size and mass growth have clearly occurred in individual systems. Parameterizing the relation between mass and size growth over $0<z<1.6$ as $R \propto M^\alpha$, we find $\alpha = 1.6 \pm 0.3$, in agreement with theoretical expectations from simulations of minor mergers. Relaxing the assumption that the velocity dispersion is unchanging, we examine growth assuming a constant ranking in galaxy velocity dispersion. This approach is applicable only to the large-dispersion tail of the distribution, but yields a consistent growth rate of $\alpha=1.4\pm0.2$. Both methods confirm that progenitor bias alone is insufficient to explain our new observations and that quiescent galaxies have grown in both size and stellar mass over $0<z<1.6$.
\end{abstract}

\keywords{galaxies: evolution --- galaxies: fundamental parameters --- galaxies: high-redshift --- galaxies: structure}


\section{Introduction}
\label{sec:intro}

Understanding the assembly history of the homogeneous population of present-day spheroidal galaxies remains an outstanding question in extragalactic astronomy. Studies of the fundamental plane of spheroidal galaxies at $z<1$ \citep{treu05, vanderwel05} confirmed that the most massive galaxies formed the bulk of their stars at $z > 2$, whereas less massive systems continued their assembly at later times. Deep near-infrared imaging meanwhile located a population of $z>2$ massive quiescent galaxies \citep{franx03}, suggesting these are the precursors of the most massive local objects. However, surprisingly, these distant red galaxies are \emph{physically small}, with half-light radii 3-5 times less than their local counterparts of similar stellar mass \citep[e.g.,][]{daddi05, trujillo06, vandokkum06, vandokkum08}. The inferred size expansion has been the source of much theoretical puzzlement, and dry mergers -- especially involving low-mass companions -- are thought to be the key growth mechanism \citep{naab09, hopkins10}.

Considerable effort has been devoted toward establishing the robustness of the relevant observations. Although stellar mass estimates are subject to uncertainties arising from assumed stellar population models, the uncertainties are thought to be insufficient to significantly change the inferred rates of growth \citep{muzzin09}. In an important step forward, \citet{newman10} inferred similar size growth rates based on more robust \emph{dynamical} mass measures over $z \simeq 0 - 1.5$. Similarly, the arrival of Wide Field Camera 3 onboard the \emph{Hubble Space Telescope} (WFC3/\HST) has allowed the light profiles of $z \simeq 2$ quiescent galaxies to be traced to many effective radii, thereby confirming the compact nature of the $z\simeq$2 sources \citep[e.g.,][]{szomoru12}, as well as providing large, homogeneous samples imaged at high spatial resolution in the rest-frame optical \citep{newman12}. 

Given the robustness of the inferred masses and sizes, the key question is the growth mechanism. While the number of observed impending mergers appears consistent with that required to account for size growth over $z \lesssim 1$, the growth rate at higher redshifts is much faster, possibly suggesting an additional mechanism \citep{newman12}. A key difficulty arises from the continual quenching of galaxies and their arrival onto the red sequence, which implies that the average size evolution for the population need not necessarily measure that of any individual galaxy. In fact, some authors have claimed a dominant role for \emph{progenitor bias} --- the later arrival of newly-quenched, potentially larger galaxies --- in interpreting size growth observations \citep[e.g.,][]{carollo13,poggianti13numberdensity}. The suggestion strikes at the heart of a fundamental problem in galaxy evolution, namely how to separate one component of a population which evolves, e.g. in size and color, over time, from a second component which joins that population at a later time. \citet{newman12} attempted to resolve this ambiguity using the evolving \emph{size distribution} and \emph{number density} of quiescent systems, arguing that the disappearance of the most compact systems could only arise from growth of
individual systems. This approach requires minimal assumptions, but it is necessarily sensitive only to the compact tail of the distribution.

A number of studies indicate that the \emph{stellar velocity dispersion} $\sigma$ of a galaxy is the most fundamental tracer of its stellar populations and halo mass \citep[e.g.,][]{graves09_I, wake12} and hence can act as a valuable identifier of a consistent population over cosmic time. In the context of size evolution, the velocity dispersion is a valuable label for several reasons. First, mergers are expected to increase the radius but change the velocity dispersion relatively little \citep[e.g.,][]{nipoti03,hopkins09scalingrel, oser12}. Second, whereas there is evidence at $z \sim 0$ for a correlation between size and stellar age at fixed mass, there is no such correlation at fixed velocity dispersion \citep[e.g.,][]{graves09_II, vanderwel09, valentinuzzi10}. This suggests that any new arrivals onto the red sequence at a given velocity dispersion do not bias the mean size of the population. Third, the number density of the highest-$\sigma$ galaxies appears to be stable over time, indicating that galaxies with $\sigma \gtrsim 280$ km~s${}^{-1}$ are in place at early times and represent a nearly fixed population \citep{bezanson12}. \citet{newman10} found no significant difference between the rates of size growth at fixed velocity dispersion and fixed mass in a preliminary sample of 17 $z \sim 1.3$ galaxies, suggesting that the role of progenitor bias in interpreting size evolution is not large.

Relatively few velocity dispersions have been measured for quiescent galaxies at high redshifts and so it has not been possible to construct the well-defined large samples necessary for constraining their number densities. For this reason, \citet{bezanson11, bezanson12} developed a photometric method to derive \emph{inferred} velocity dispersions for 5000 quiescent galaxies over $0<z<1.3$ within the Newfirm Medium Band Survey \citep[NMBS,][]{whitaker11}. This approach uses the stellar masses, effective radii, and S\'{e}rsic indices of the distant sample to estimate velocity dispersions using a formula calibrated locally using SDSS data \citep{taylor10}. However, given this calibration may well evolve with redshift, direct spectroscopic measurements remain indispensable. Although the current spectroscopic datasets at $z \gtrsim 1$ appear consistent with the locally-derived calibration \citep[e.g.,][]{vandesande13}, the sample sizes are too small for this approach to be robust.

A further benefit of securing velocity dispersions from spectroscopic data is the ability to compare the relationship between stellar and dynamical masses for individual objects. The ratio of stellar to dynamical mass, $\Mstar/\Mdyn$, is a potentially valuable tracer of the likely mechanism by which galaxies grow \citep[e.g.,][]{hopkins09scalingrel,hilz13}. Specifically, under merger-driven growth, the ratio measured within the effective radius should decrease with time. This decrease would be stronger in the case of minor mergers. Some tentative support for this suggestion was discussed by \citet{vandesande13} using a sample of 5 galaxies with $1.5<z<2.1$.

To address the above issues, and building on earlier work by \citet{newman10}, we have completed a new spectroscopic survey of over 100 $z>1$ massive galaxies utilizing the red-sensitive CCD installed in the Keck LRIS spectrograph, thus providing nearly a four-fold increase in the sample size over earlier work. Such a large sample allows us to examine size growth at fixed velocity dispersion, thereby addressing the question of progenitor bias, as well as the relationship between stellar and dynamical mass over $0<z<1.5$. Later papers in this series will further address the issues of progenitor bias via spectroscopic indicators of recently-quenched galaxies  (S. Belli et al in prep).

A plan of the paper follows. In Section \ref{sec:data} we describe the selection of the Keck LRIS sample, the spectroscopic observations and their data reduction. We also discuss the auxiliary data used for deriving sizes and stellar masses, as well as the comparison sample of local galaxies; and we present the selection of quiescent galaxies based on rest-frame colors. In Section \ref{sec:measurements} we derive the key physical properties: size, stellar mass and stellar velocity dispersion, essential for our analysis, and we discuss the relevant uncertainties. In Section \ref{sec:dynmasses} we calculate the dynamical masses and discuss the stellar-dynamical mass relation and its redshift evolution. In Section \ref{sec:size_evo} we investigate the size growth of quiescent galaxies using the stellar velocity dispersion as a tracer of populations connected over cosmic time. Finally, we summarize our main results and discuss their implications in Section \ref{sec:discussion}.
Throughout this work we use AB magnitudes, and assume a $\Lambda$CDM cosmology with $\Omega_m = 0.3$, $\Omega_\Lambda = 0.7$, and $H_0 = 70 $ km s$^{-1}$ Mpc$^{-1}$.


\section{Data}
\label{sec:data}

\subsection{Sample Selection}

We selected spectroscopic targets from various photometric catalogs in three well-studied fields: COSMOS, GOODS-South and EGS. The public photometric data are described in Section \ref{sec:aux_data} and Appendix \ref{appendix:photometry}. Galaxies were selected with photometric redshifts in the range $ 0.9 < z_\mathrm{phot} < 1.6 $ and stellar masses (calculated from broad-band photometry, see Section \ref{sec:sed_fit}) larger than $10^{10.6} \Msun$. In designing slitmasks we gave priority to massive and red objects according to their rest-frame spectral energy distributions, and added extra sources from a less-strictly selected sample. In the first observing run (see Section \ref{sec:observations} and Table \ref{tab:masks}) we used slightly different criteria: $1 < z_\mathrm{phot} < 2$, magnitude in the $z$ band brighter than 23.5, and spheroidal morphology in \HST\ ACS imaging. Objects brighter than $K \sim 22$ were used as additional sources. To this sample, we added 17 galaxies published by \citet{newman10} also observed with LRIS 
in the EGS, GOODS-North and SSA22 fields. These objects were selected to have $I<23.5$, $I-K_S > 2$, and a spheroidal morphology.

\begin{deluxetable}{lccc}
\tabletypesize{\footnotesize}
\tablewidth{0pc}
\tablecaption{LRIS Observations \label{tab:masks}}
\tablehead{
\colhead{Slitmask} & \colhead{Run\tablenotemark{a}} & \colhead{Seeing (arcsec)} & \colhead{Exp. Time (min)} }
\startdata
GOODS-S 1	& A		& 0.8		& 420		\\
COSMOS 1	& A		& 1.0		& 420		\\
COSMOS 2	& B,C		& 0.8		& 360		\\
COSMOS 3	& C		& 0.9		& 240		\\
COSMOS 4	& D		& 1.0		& 220		\\
EGS 1		& D		& 1.0		& 260		\\
EGS 2		& D		& 1.6		& 180	
\enddata
\tablenotetext{a}{Observing runs: A: 2011 January 6--9; B: 2011 November 21, 22; C: 2012 January 22, 23; D: 2012 April 17,18.}
\end{deluxetable}

\subsection{Spectroscopic Observations and Data Reduction}
\label{sec:observations}

We observed the selected galaxies using the upgraded red arm \citep{rockosi10} of the Low-Resolution Imaging Spectrograph \citep[LRIS,][]{oke95} on the Keck I telescope. We used 1\arcsec\ wide slits and the 600 mm$^{-1}$ grating blazed at 1 $\mu$m, with a resulting velocity resolution of $\sigma_{\rm instr} \sim 60$ km s$^{-1}$ at 9000 \AA. The spectra were taken over four observing runs in 2011 and 2012. We observed a total of seven slitmasks targeting 20-25 objects each, listed in Table \ref{tab:masks}. The total integration times varied from 3 to 7 hours per mask, with individual frames having a typical exposure time of 1200 seconds.

\begin{figure}[tbp]
\centering
\includegraphics[width=0.45\textwidth]{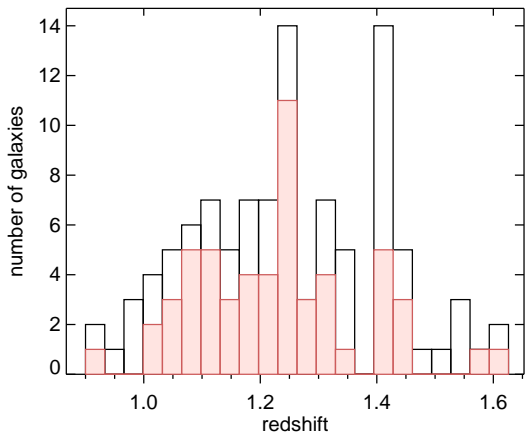}
\caption{Redshift distribution of the spectroscopic sample of 103 galaxies, including the sample published by \citet{newman10}. The subsample of 56 quiescent galaxies used in the subsequent analysis (see Section \ref{subsec:redselection}) is shown in red.}
\label{fig:n_z}
\end{figure}
	
The data were reduced using a pipeline based on the code developed by \citet{kelson03}. Each frame was corrected for bias and flat-fielded, and the sky emission was modeled and subtracted. The 1D spectra were then optimally extracted from the stacked frames using weights derived by fitting a Gaussian to the spatial profile of each trace. The sky spectrum was also extracted from each slit with the purpose of accurately measuring the instrumental resolution. Telluric corrections and flux calibrations were determined using observations of standard white dwarfs. To ensure a good telluric calibration, the spectra of the standard stars were broadened to match the resolution of the science observations, where necessary.

From these slitmasks we obtained 86 spectra with at least one clear feature that allows us to determine the spectroscopic redshift. To this sample we add 17 galaxies from \citet{newman10} which were observed with LRIS with slightly longer exposure times and which were reduced in a similar way. The redshift distribution of the full spectroscopic sample of 103 sources is shown in Figure \ref{fig:n_z}. Modest overdensities are apparent at $z\sim1.25$ and $z\sim1.4$. 

\begin{figure*}[tbp]
    
\begin{minipage}{\textwidth}
   \centering
 \raisebox{-0.5\height}{\includegraphics[width=0.48\textwidth]{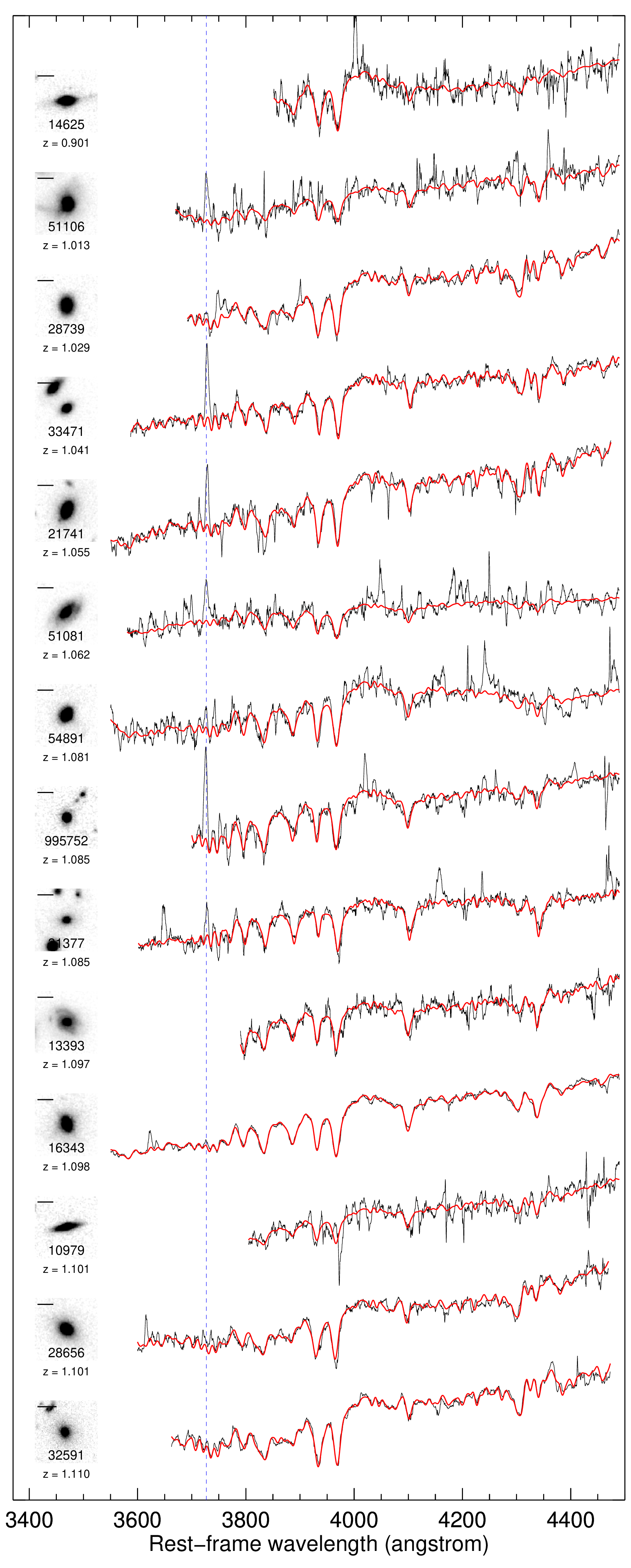}}
   \hspace*{0.02\textwidth}
 \raisebox{-0.5\height}{\includegraphics[width=0.48\textwidth]{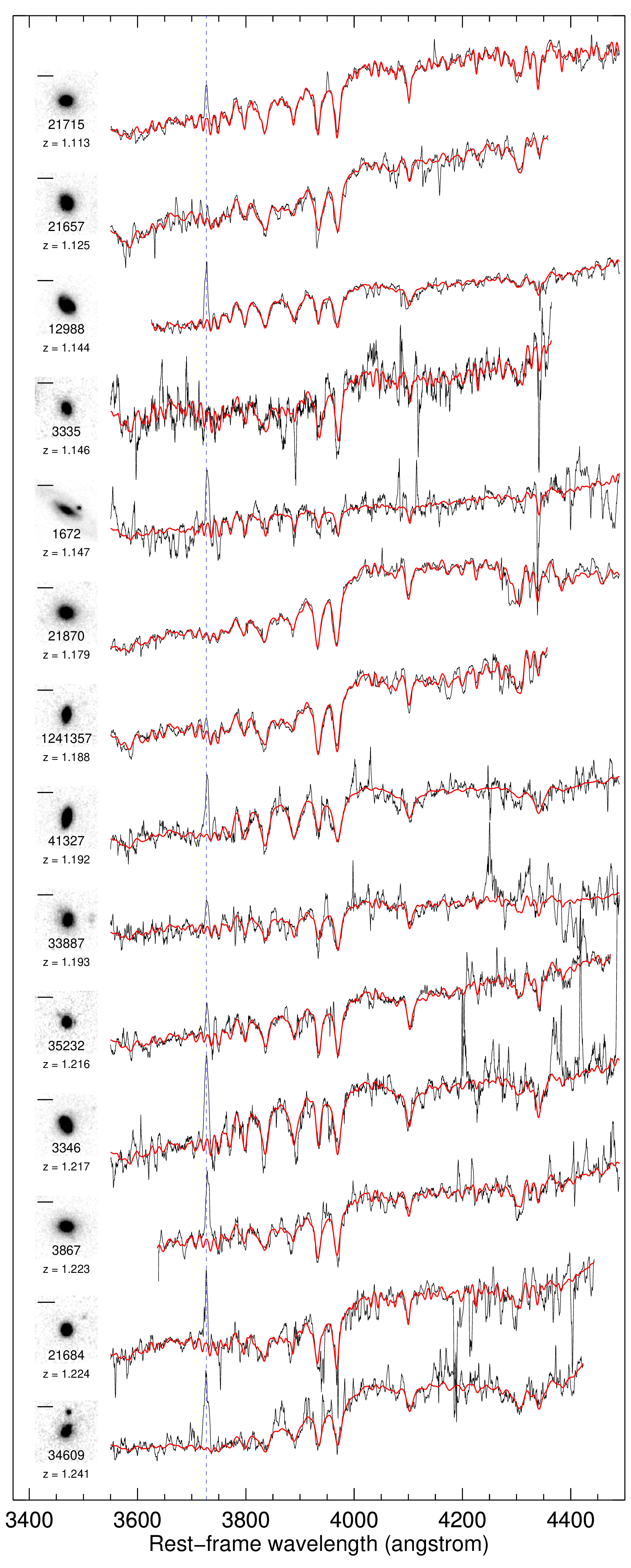}}
\end{minipage}

\caption{Observed LRIS spectra of the 56 quiescent galaxies for which accurate velocity dispersions were measured, sorted by redshift. The spectra are inverse-variance smoothed with a window of 21 pixels, corresponding to $\sim 7.5$ \AA\ in the rest-frame (16.8 \AA\ in the observed frame). The vertical blue dashed line is the expected position for the \OII\ emission line. For each galaxy, the \HST\ cutout (with a 10 kpc ruler), the ID and the spectroscopic redshift are shown on the left, and the best-fit spectrum is overplotted in red. The \HST\ images are in the F160W band except for the objects 32591 and 37085, for which we use F814W.}
\label{fig:sfit}
\end{figure*}

\begin{figure*}[tbp]
    
\begin{minipage}{\textwidth}
   \centering
 \raisebox{-0.5\height}{\includegraphics[width=0.48\textwidth]{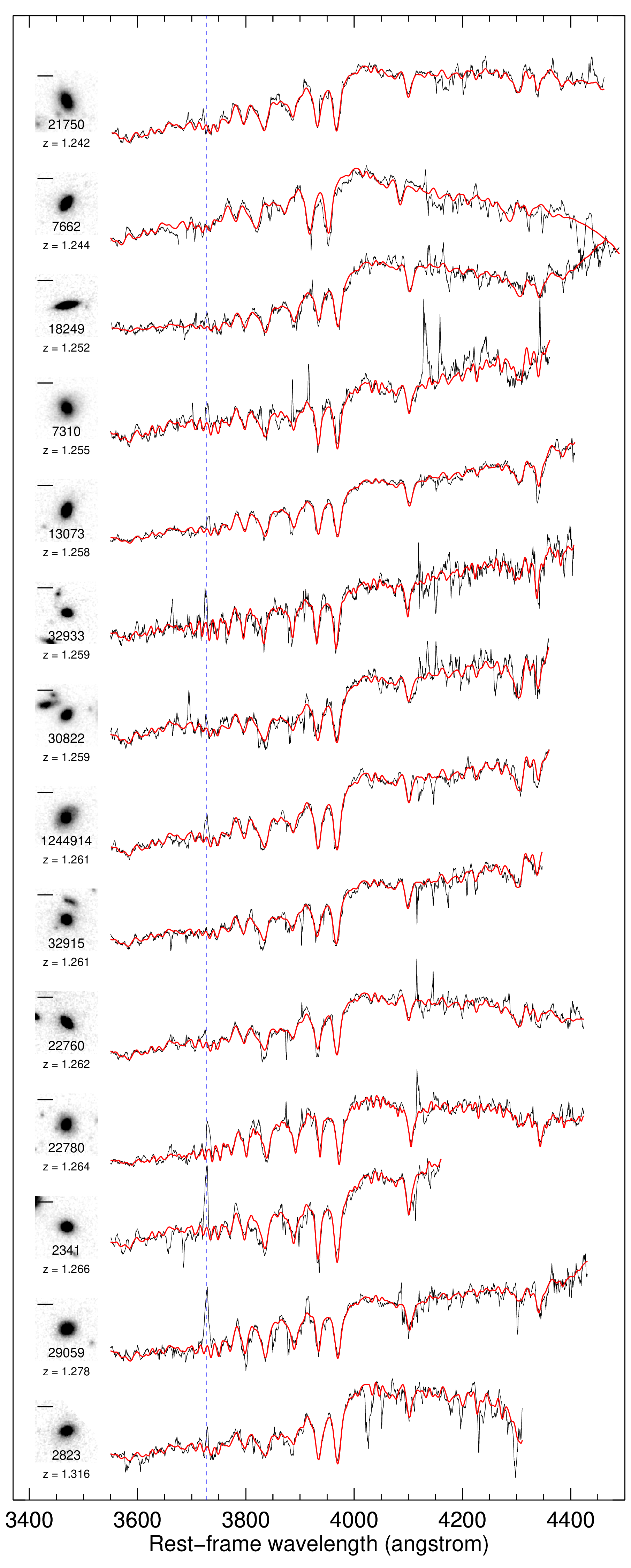}}
   \hspace*{0.02\textwidth}
 \raisebox{-0.5\height}{\includegraphics[width=0.48\textwidth]{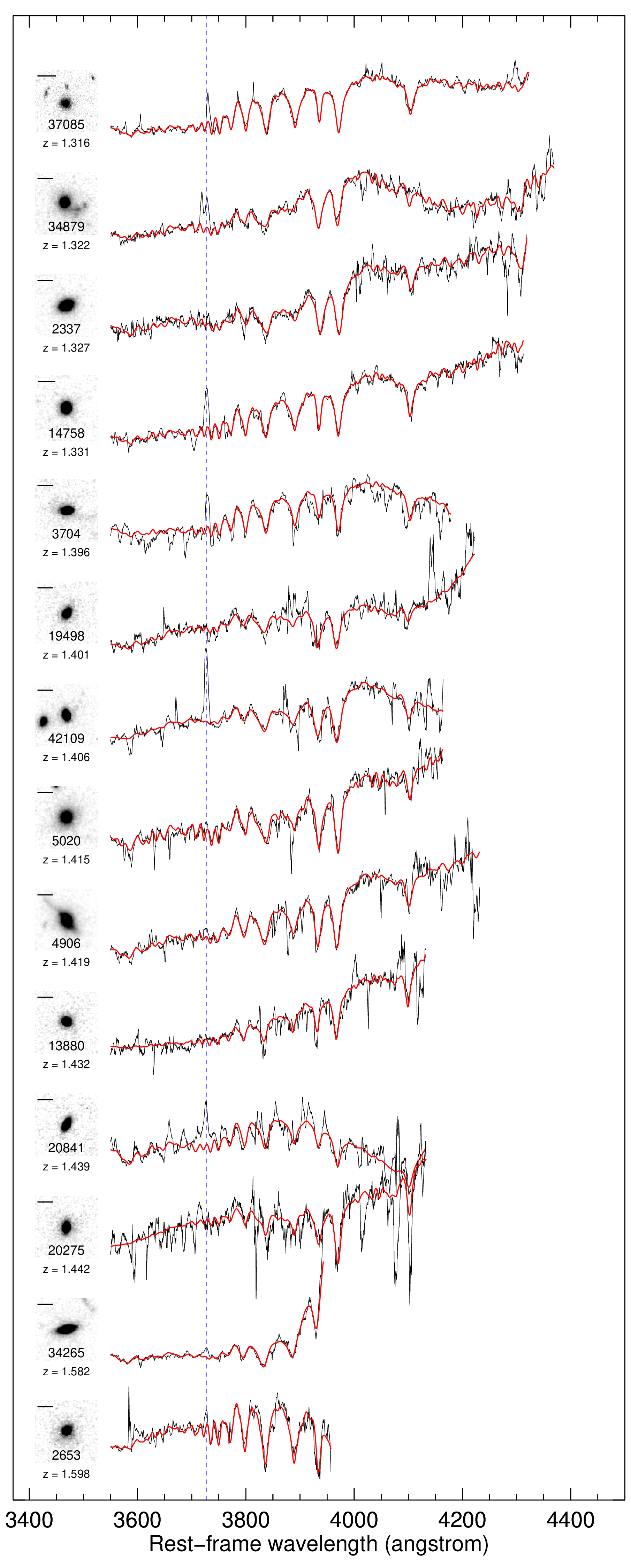}}
\end{minipage}

\addtocounter{figure}{-1}
\caption{Continued.}
\label{fig:sfit2}
\end{figure*}

\subsection{Auxiliary Data}
\label{sec:aux_data}

In order to measure stellar masses and other properties we use photometric data for the galaxies in our sample from a number of publicly available catalogs. Space and ground-based observations from the near-UV to the near-infrared are available for every object. All except three galaxies also have \emph{Spitzer} IRAC public data. In Appendix \ref{appendix:photometry} we describe in some detail the compilation of photometric data.

Space-based optical and near-infrared observations are critical for an accurate estimate of the size of high-redshift galaxies, and we exclusively use publicly available \HST\ data for surface brightness fitting. For the GOODS-N, EGS, GOODS-S, and COSMOS fields we used the F160W data from the Cosmic Assembly Near-IR Deep Extragalactic Legacy Survey \citep[CANDELS,][]{grogin11, koekemoer11}. Since most of our GOODS-S sample is outside the area probed by the CANDELS observations, we also used F160W data from the Early Release Science survey \citep{windhorst11}. For the three objects in SSA22, near-infrared \HST\ observations are not publicly available, and we use the F814W data presented in \citet{newman10}.

\subsection{SDSS Data}

We have selected a sample of galaxies from the Sloan Digital Sky Survey \citep[SDSS DR7,][]{abazajian09} that will be useful for comparing the properties of our high-redshift sample with the population of local galaxies. We make use of the NYU Value Added Catalog \citep{blanton05}, which includes many derived properties. SDSS galaxies were selected with spectroscopic redshifts within the interval $0.05<z<0.08$, excluding objects flagged as hosting an active galactic nucleus (AGN) according to the flux ratios of emission lines. We also discarded galaxies with poor spectral fits, and those with a very large uncertainty on the measured velocity dispersion.

For each selected galaxy we use the NYU catalog determination of its \Sersic\ index and the effective radius obtained by a \Sersic\ profile fit to the $r$ band imaging \citep{Blanton05sersic}, the velocity dispersion measured from the optical spectrum, and the SDSS and 2MASS photometry. Since we require $J$ band photometry for selecting quiescent galaxies in a manner similar to that adopted at high redshift (see Section \ref{subsec:redselection}), we only consider the subsample of galaxies detected in the 2MASS imaging survey. This survey is shallower than the SDSS, but this is not an issue for our study, since above $10^{10.6} \Msun$ (which is the limiting mass for the high-redshift sample) more than 95\% of the SDSS galaxies are detected in $J$. This selection gives a sample of 68738 objects. Finally, we match each object to the MPA-JHU catalog \citep{kauffmann03}, from which we take stellar masses and star formation rates, which are calculated from the broad-band photometry assuming a \citet{chabrier03} IMF.

\subsection{Selecting Quiescent Galaxies}
\label{subsec:redselection}

\begin{figure}[tbp]
\centering
\includegraphics[width=0.45\textwidth]{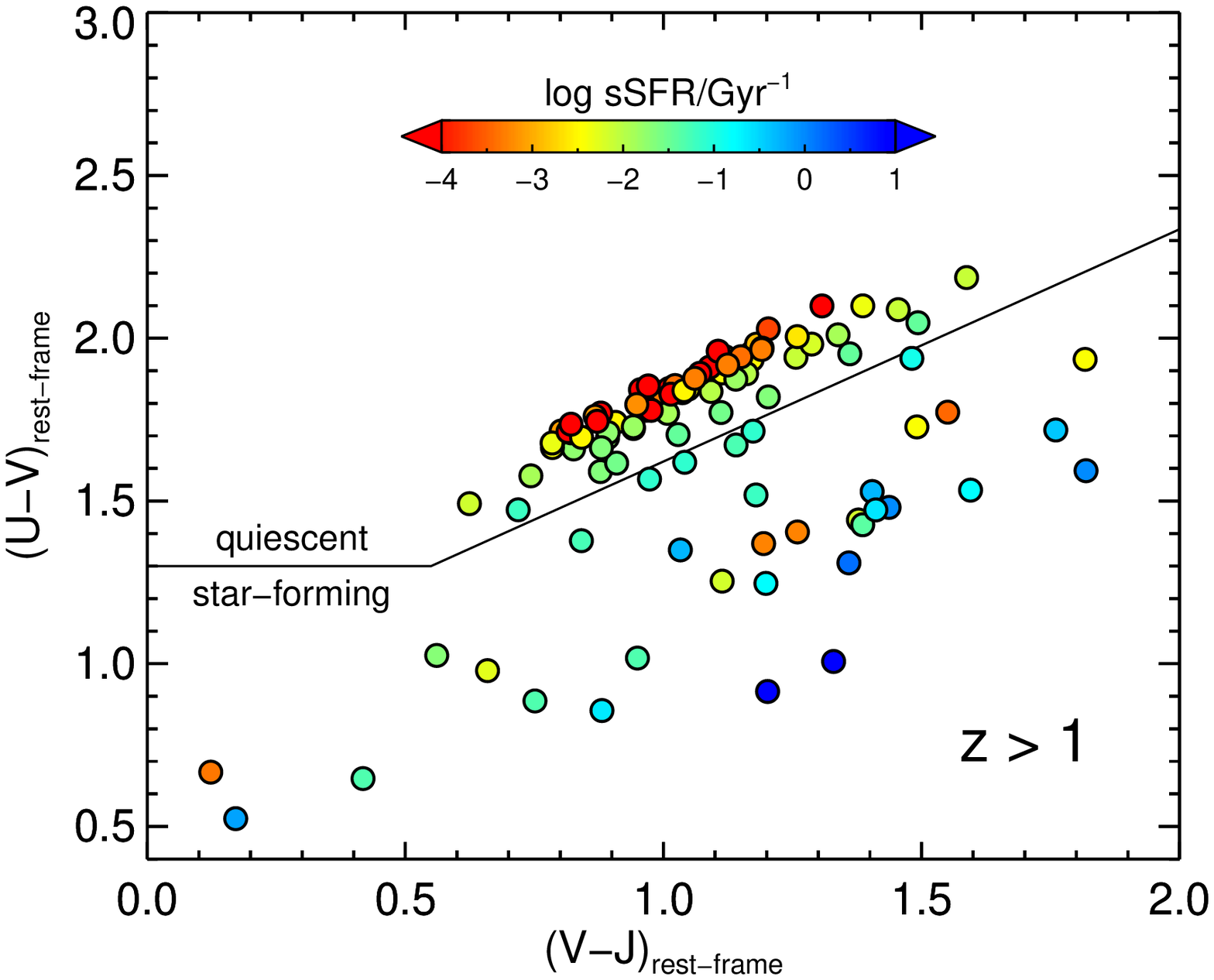}
\includegraphics[width=0.45\textwidth]{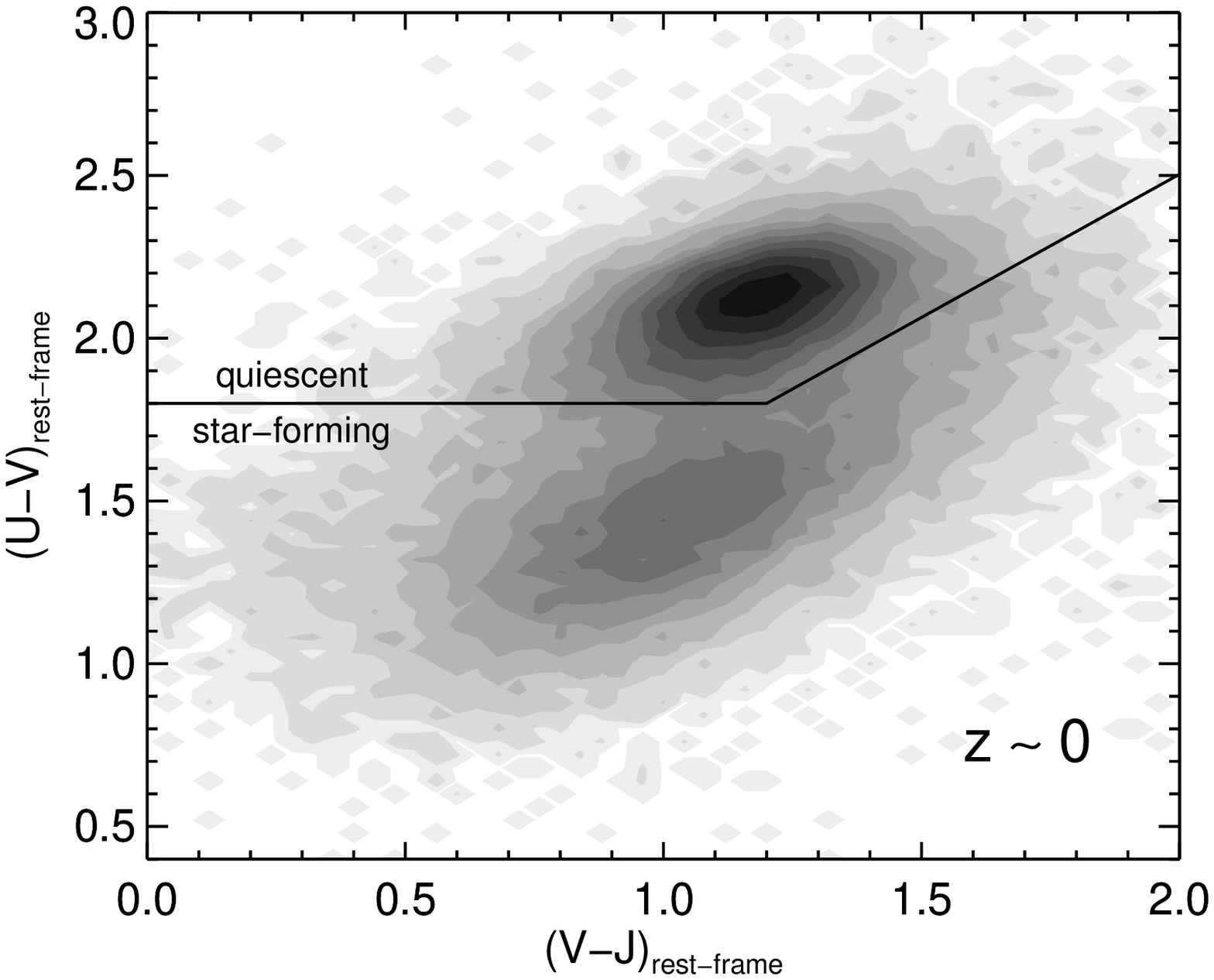}
\caption{Selection of the quiescent subsample. \emph{Top:} the total sample of 103 high-redshift galaxies in the $UVJ$ plane. Colors denote specific star formation rates measured from SED fitting. \emph{Bottom:} $UVJ$ diagram for the SDSS comparison sample. In each panel, the solid line marks the adopted division between quiescent and star-forming galaxies.}
\label{fig:redselection}
\end{figure}

The main goal of this work is to study the evolution of quiescent galaxies. To identify this type of galaxy we primarily rely a color-color selection. Rest-frame $UVJ$ magnitudes are determined by integrating the synthetic spectrum that best fits the observed SED (see Section \ref{sec:sed_fit}). The $U-V$ versus $V-J$ plane is shown in the top panel of Figure \ref{fig:redselection}. In this plane quiescent galaxies tend to form a tight sequence distinctly separated from the region occupied by star-forming galaxies \citep[e.g.][]{williams09}. In the figure, the SED-derived specific star formation rates (sSFR, star formation rate per unit stellar mass, see Section \ref{sec:sed_fit}) are shown for each object using a color code. The red sequence is clearly visible and composed of galaxies with low sSFR, roughly less than 0.1 Gyr$^{-1}$. An appropriate division between red, passive galaxies and blue, star-forming ones is shown \citep[black line, see also][]{whitaker11}. 

Out of the total of 103 objects, this color-color selection yields 69 quiescent galaxies, of which 56 have excellent quality, high signal-to-noise spectra (see Section \ref{sec:dispersions}); these form the primary sample for analysis in this paper. Their redshift distribution is shown via the shaded histogram in Figure \ref{fig:n_z}, and their properties are summarized in Table \ref{tab:sample}. The observed spectra, together with \HST\ image cutouts, are shown in Figure \ref{fig:sfit}. The rest-frame coverage is roughly centered on 4000 \AA, but changes with redshift and slit position on the mask. The Ca II H and K absorption lines are well detected for all the objects except those at $z>1.5$, while Balmer absorption lines vary from very strong to almost absent. A detailed spectroscopic study of the total sample, including the subset of quiescent galaxies, will be presented in a future work (S. Belli et al., in preparation).

The \OII\ emission line is clearly visible in a number of spectra, and could be due either to some residual star formation or to low-ionization nuclear emission-line region (LINER) activity. Out of the 53 objects for which the line falls in the observed wavelength range, 25 present clear [OII] emission, with an equivalent width larger than 3 \AA. We calculate an average equivalent width of 8 \AA\ (and never exceeding 15 \AA), and use the calibration of \citet{kewley04} to derive a rough estimate of the star formation rate. We obtain a mean value of specific star formation rate of 0.032 Gyr$^{-1}$, and a maximum of 0.12 Gyr$^{-1}$. These values are consistent with or larger than the ones resulting from the SED fitting (see Figure \ref{fig:redselection} and Section \ref{sec:sed_fit}). We note that this method does not take into account the LINER contribution, which is expected to be important for this type of galaxies at $z\sim0$ \citep{yan06} as well as $z\sim1$ \citep{lemaux10}, and therefore yields only an upper limit on the star formation rate. We checked that a more strict selection of quiescent galaxies that excludes objects with detected [OII] emission at both high and low redshift does not significantly change our results. Finally, we find that 3 out of 56 objects are detected in publicly available X-Ray data. These galaxies are likely to host AGN activity, and we list them in Table \ref{tab:sample}.

Using the same criteria as for the high-redshift galaxies, we selected a sample of quiescent galaxies from the SDSS population for comparison purposes. We use InterRest \citep{taylor09} to calculate the rest-frame colors from the observed SDSS and 2MASS photometry, and we show the $UVJ$ diagram for local galaxies in the bottom panel of Figure \ref{fig:redselection}. In this case, galaxies present a clear bimodal distribution, even though the red sequence is shifted toward redder colors. Adopting the definition for the quiescent sample shown in the plot, we obtain 37852 objects.


\section{Physical properties of the sample}
\label{sec:measurements}

In this section we derive the physical properties of the sample of high-redshift galaxies using photometric, imaging and spectroscopic data.

\subsection{Size Measurement}
\label{sec:sizes}

To study the size and structure of each galaxy we make use of the \HST\ F160W data, which correspond to a rest-frame wavelength in the $R$ or $I$ band depending on the redshift. For two objects only F814W (rest-frame UV) data are available.  We fit a 2D \citet{sersic63} profile to the surface brightness of every galaxy using the GALFIT code \citep{peng02}. Adjacent objects were identified from the SExtractor \citep{bertin96} segmentation map and masked out or, when bright and close enough to influence the central region of the object, fit simultaneously. Point spread functions were derived from isolated bright stars.

The output parameters from the fitting procedure include the total flux, the \Sersic\ index $n$, the axis ratio $q$, and the circularized effective radius $\R = a \sqrt{q} $, where $a$ is the effective (i.e., half-light) semi-major axis, and are listed in Table \ref{tab:sample}. We adopt a $10\%$ uncertainty on all the size measurements, in agreement with the tests performed by \citet{newman12} who used similar data and procedures and whose estimates are consistent with other studies \citep[e.g.,][]{vanderwel08}.


\begin{deluxetable*}{llcccccccccc}
\tabletypesize{\footnotesize}
\tablewidth{0pc}
\tablecaption{Physical Properties of the Sample of Quiescent Galaxies \label{tab:sample}}
\tablehead{
\colhead{Object ID} & \colhead{Slitmask} & \colhead{R.A.} & \colhead{Decl.} & \colhead{$z$} & \colhead{\sigmae} & \colhead{\R} & \colhead{$n$} & \colhead{$q$} & \colhead{$\log \Mstar/\Msun$} & \colhead{$\log \Mdyn/\Msun$}
\\
 &  & (J2000) & (J2000)  & & (km s$^{-1}$) & (kpc) & & & & 
} 
\startdata
14625\tablenotemark{b} & COSMOS 3 & 150.15839 & 2.4154 & 0.901 & $ 283 \pm 49 $ & 2.40 & 2.1 & 0.40 & $ 10.72 \pm 0.09 $ & $ 11.35 \pm 0.16 $  \\
51106 & EGS 2 & 214.92057 & 52.8659 & 1.013 & $ 252 \pm 37 $ & 5.99 & 5.3 & 0.78 & $ 11.26 \pm 0.10 $ & $ 11.65 \pm 0.14 $  \\
28739 & EGS 1 & 214.62827 & 52.7157 & 1.029 & $ 238 \pm 11 $ & 1.98 & 3.5 & 0.70 & $ 10.94 \pm 0.08 $ & $ 11.11 \pm 0.06 $  \\
33471 & COSMOS 3 & 150.15414 & 2.4157 & 1.041 & $ 176 \pm 12 $ & 1.50 & 2.6 & 0.66 & $ 10.60 \pm 0.02 $ & $ 10.73 \pm 0.07 $  \\
21741 & N10 (EGS)  & 214.98510 & 52.9512 & 1.055 & $ 211 \pm 14 $ & 2.28 & 2.2 & 0.53 & $ 10.93 \pm 0.10 $ & $ 11.07 \pm 0.07 $  \\
51081 & EGS 2 & 214.90243 & 52.8637 & 1.062 & $ 233 \pm 58 $ & 4.36 & 2.2 & 0.53 & $ 10.89 \pm 0.11 $ & $ 11.44 \pm 0.22 $  \\
54891 & EGS 2 & 214.92122 & 52.8878 & 1.081 & $ 232 \pm 37 $ & 1.27 & 3.2 & 0.73 & $ 10.72 \pm 0.06 $ & $ 10.90 \pm 0.14 $  \\
995752 & COSMOS 1 & 150.16466 & 2.2783 & 1.085 & $ 199 \pm 44 $ & 1.20 & 4.0 & 0.78 & $ 10.27 \pm 0.05 $ & $ 10.74 \pm 0.20 $  \\
31377 & COSMOS 4 & 150.05580 & 2.2718 & 1.085 & $ 133 \pm 18 $ & 4.88 & 4.9 & 0.63 & $ 10.83 \pm 0.09 $ & $ 11.00 \pm 0.13 $  \\
13393 & COSMOS 3 & 150.06162 & 2.3881 & 1.097 & $ 175 \pm 21 $ & 7.18 & 3.5 & 0.80 & $ 11.16 \pm 0.05 $ & $ 11.41 \pm 0.12 $  \\
16343 & COSMOS 3 & 150.09793 & 2.4468 & 1.098 & $ 290 \pm 8 $ & 1.95 & 8.0 & 0.65 & $ 11.04 \pm 0.03 $ & $ 11.28 \pm 0.05 $  \\
10979 & COSMOS 3 & 150.16008 & 2.3488 & 1.101 & $ 213 \pm 116 $ & 2.02 & 1.9 & 0.18 & $ 10.66 \pm 0.08 $ & $ 11.03 \pm 0.47 $  \\
28656 & EGS 1 & 214.67508 & 52.7163 & 1.101 & $ 251 \pm 15 $ & 2.77 & 5.4 & 0.70 & $ 11.08 \pm 0.08 $ & $ 11.31 \pm 0.07 $  \\
32591\tablenotemark{a} & N10 (SSA22)  & 334.35290 & 0.2734 & 1.110 & $ 245 \pm 10 $ & 4.40 & 2.4 & 0.86 & $ 11.22 \pm 0.11 $ & $ 11.49 \pm 0.06 $  \\
21715\tablenotemark{b} & N10 (EGS)  & 214.97000 & 52.9910 & 1.113 & $ 109 \pm 8 $ & 1.99 & 4.0 & 0.77 & $ 10.83 \pm 0.07 $ & $ 10.44 \pm 0.08 $  \\
21657 & N10 (EGS)  & 215.00590 & 52.9754 & 1.125 & $ 270 \pm 13 $ & 2.14 & 2.5 & 0.74 & $ 10.97 \pm 0.09 $ & $ 11.26 \pm 0.06 $  \\
12988 & COSMOS 3 & 150.11500 & 2.3810 & 1.144 & $ 183 \pm 16 $ & 2.70 & 3.1 & 0.84 & $ 10.94 \pm 0.05 $ & $ 11.02 \pm 0.09 $  \\
3335 & COSMOS 4 & 150.11756 & 2.2226 & 1.146 & $ 121 \pm 19 $ & 1.33 & 5.1 & 0.61 & $ 10.67 \pm 0.04 $ & $ 10.35 \pm 0.14 $  \\
1672 & COSMOS 4 & 150.11025 & 2.1940 & 1.147 & $ 131 \pm 37 $ & 5.83 & 1.9 & 0.35 & $ 11.04 \pm 0.05 $ & $ 11.07 \pm 0.25 $  \\
21870 & N10 (EGS)  & 214.98450 & 52.9613 & 1.179 & $ 230 \pm 12 $ & 3.36 & 5.5 & 0.80 & $ 11.02 \pm 0.07 $ & $ 11.32 \pm 0.06 $  \\
1241357 & COSMOS 1 & 150.11053 & 2.3235 & 1.188 & $ 207 \pm 13 $ & 1.06 & 5.0 & 0.43 & $ 10.86 \pm 0.04 $ & $ 10.72 \pm 0.07 $  \\
41327 & EGS 2 & 214.86345 & 52.8040 & 1.192 & $ 324 \pm 41 $ & 1.17 & 2.9 & 0.35 & $ 10.80 \pm 0.05 $ & $ 11.16 \pm 0.12 $  \\
33887 & EGS 1 & 214.77293 & 52.7556 & 1.193 & $ 162 \pm 33 $ & 3.76 & 2.1 & 0.75 & $ 10.74 \pm 0.11 $ & $ 11.06 \pm 0.18 $  \\
35232 & EGS 1 & 214.73653 & 52.7618 & 1.216 & $ 191 \pm 19 $ & 1.01 & 3.6 & 0.77 & $ 10.56 \pm 0.04 $ & $ 10.63 \pm 0.10 $  \\
3346 & COSMOS 4 & 150.11237 & 2.2223 & 1.217 & $ 185 \pm 22 $ & 2.62 & 1.2 & 0.64 & $ 10.81 \pm 0.05 $ & $ 11.02 \pm 0.11 $  \\
3867 & GOODS-S 1 & 53.10946 & -27.7641 & 1.223 & $ 184 \pm 27 $ & 2.74 & 6.7 & 0.72 & $ 10.67 \pm 0.08 $ & $ 11.03 \pm 0.14 $  \\
21684 & N10 (EGS)  & 214.98130 & 52.9500 & 1.224 & $ 131 \pm 23 $ & 0.95 & 2.2 & 0.80 & $ 10.55 \pm 0.08 $ & $ 10.28 \pm 0.16 $  \\
34609 & COSMOS 2 & 150.16114 & 2.5049 & 1.241 & $ 279 \pm 159 $ & 6.51 & 8.0 & 0.67 & $ 11.04 \pm 0.08 $ & $ 11.77 \pm 0.50 $  \\
21750 & N10 (EGS)  & 215.03490 & 52.9829 & 1.242 & $ 264 \pm 16 $ & 2.59 & 5.2 & 0.57 & $ 11.03 \pm 0.07 $ & $ 11.32 \pm 0.07 $  \\
7662 & N10 (GOODS-N)  & 189.26810 & 62.2264 & 1.244 & $ 293 \pm 37 $ & 0.98 & 3.1 & 0.35 & $ 10.92 \pm 0.07 $ & $ 10.99 \pm 0.12 $  \\
18249 & COSMOS 2 & 150.10303 & 2.4821 & 1.252 & $ 286 \pm 109 $ & 1.60 & 1.1 & 0.16 & $ 10.77 \pm 0.04 $ & $ 11.18 \pm 0.33 $  \\
7310 & COSMOS 4 & 150.05791 & 2.2904 & 1.255 & $ 176 \pm 16 $ & 4.34 & 3.8 & 0.87 & $ 11.13 \pm 0.07 $ & $ 11.19 \pm 0.09 $  \\
13073 & COSMOS 3 & 150.12479 & 2.3823 & 1.258 & $ 265 \pm 12 $ & 1.20 & 2.8 & 0.50 & $ 10.97 \pm 0.03 $ & $ 10.99 \pm 0.06 $  \\
32933 & COSMOS 3 & 150.09624 & 2.3770 & 1.259 & $ 131 \pm 19 $ & 0.91 & 2.5 & 0.56 & $ 10.50 \pm 0.05 $ & $ 10.26 \pm 0.13 $  \\
30822 & COSMOS 4 & 150.09089 & 2.2252 & 1.259 & $ 271 \pm 25 $ & 1.82 & 2.5 & 0.68 & $ 10.96 \pm 0.07 $ & $ 11.19 \pm 0.09 $  \\
1244914 & COSMOS 1 & 150.17400 & 2.3010 & 1.261 & $ 252 \pm 13 $ & 4.99 & 5.5 & 0.79 & $ 11.18 \pm 0.07 $ & $ 11.57 \pm 0.06 $  \\
32915 & COSMOS 3 & 150.14620 & 2.3743 & 1.261 & $ 264 \pm 17 $ & 1.33 & 6.3 & 0.82 & $ 10.88 \pm 0.05 $ & $ 11.03 \pm 0.07 $  \\
22760 & N10 (EGS)  & 215.13690 & 53.0172 & 1.262 & $ 232 \pm 17 $ & 0.94 & 2.4 & 0.37 & $ 10.83 \pm 0.06 $ & $ 10.77 \pm 0.08 $  \\
22780 & N10 (EGS)  & 215.13170 & 53.0162 & 1.264 & $ 88 \pm 18 $ & 2.28 & 4.2 & 0.77 & $ 10.75 \pm 0.07 $ & $ 10.31 \pm 0.18 $  \\
2341 & N10 (GOODS-N)  & 189.06340 & 62.1623 & 1.266 & $ 190 \pm 27 $ & 1.21 & 3.8 & 0.71 & $ 10.87 \pm 0.06 $ & $ 10.70 \pm 0.13 $  \\
29059 & EGS 1 & 214.61016 & 52.7188 & 1.278 & $ 208 \pm 16 $ & 1.62 & 4.3 & 0.77 & $ 10.90 \pm 0.06 $ & $ 10.91 \pm 0.08 $  \\
2823 & N10 (GOODS-N)  & 188.93450 & 62.2068 & 1.316 & $ 215 \pm 21 $ & 3.26 & 5.4 & 0.64 & $ 11.01 \pm 0.16 $ & $ 11.24 \pm 0.10 $  \\
37085\tablenotemark{a} & N10 (SSA22)  & 334.35020 & 0.3032 & 1.316 & $ 164 \pm 14 $ & 2.51 & 1.8 & 0.94 & $ 10.60 \pm 0.15 $ & $ 10.89 \pm 0.09 $  \\
34879 & COSMOS 2 & 150.13138 & 2.5238 & 1.322 & $ 213 \pm 53 $ & 5.45 & 8.0 & 0.87 & $ 11.23 \pm 0.05 $ & $ 11.46 \pm 0.22 $  \\
2337 & COSMOS 4 & 150.10076 & 2.2058 & 1.327 & $ 279 \pm 20 $ & 1.54 & 3.5 & 0.70 & $ 11.04 \pm 0.06 $ & $ 11.14 \pm 0.08 $  \\
14758\tablenotemark{b}  & COSMOS 3 & 150.06416 & 2.4179 & 1.331 & $ 156 \pm 16 $ & 0.83 & 2.2 & 0.84 & $ 10.71 \pm 0.03 $ & $ 10.37 \pm 0.10 $  \\
3704 & N10 (GOODS-N)  & 189.11320 & 62.1325 & 1.396 & $ 191 \pm 23 $ & 0.98 & 4.1 & 0.42 & $ 10.47 \pm 0.06 $ & $ 10.62 \pm 0.11 $  \\
19498 & COSMOS 2 & 150.11063 & 2.5038 & 1.401 & $ 250 \pm 39 $ & 0.84 & 4.2 & 0.46 & $ 10.75 \pm 0.07 $ & $ 10.79 \pm 0.14 $  \\
42109 & N10 (EGS)  & 215.12170 & 52.9575 & 1.406 & $ 369 \pm 48 $ & 0.73 & 2.3 & 0.41 & $ 10.77 \pm 0.07 $ & $ 11.06 \pm 0.12 $  \\
5020 & GOODS-S 1 & 53.17976 & -27.7116 & 1.415 & $ 181 \pm 54 $ & 2.07 & 4.6 & 0.88 & $ 10.83 \pm 0.08 $ & $ 10.90 \pm 0.26 $  \\
4906 & GOODS-S 1 & 53.18302 & -27.7090 & 1.419 & $ 298 \pm 26 $ & 2.33 & 3.7 & 0.59 & $ 11.34 \pm 0.07 $ & $ 11.38 \pm 0.09 $  \\
13880 & COSMOS 3 & 150.07210 & 2.4001 & 1.432 & $ 169 \pm 70 $ & 0.87 & 2.6 & 0.62 & $ 10.64 \pm 0.07 $ & $ 10.46 \pm 0.36 $  \\
20841 & COSMOS 2 & 150.17009 & 2.5256 & 1.439 & $ 267 \pm 52 $ & 1.43 & 1.3 & 0.35 & $ 10.65 \pm 0.06 $ & $ 11.07 \pm 0.18 $  \\
20275 & COSMOS 2 & 150.07093 & 2.5164 & 1.442 & $ 221 \pm 70 $ & 1.36 & 4.0 & 0.51 & $ 10.80 \pm 0.07 $ & $ 10.89 \pm 0.28 $  \\
34265 & COSMOS 2 & 150.17016 & 2.4811 & 1.582 & $ 377 \pm 54 $ & 0.92 & 2.9 & 0.22 & $ 11.33 \pm 0.04 $ & $ 11.18 \pm 0.13 $  \\
2653 & N10 (GOODS-N)  & 188.96250 & 62.2286 & 1.598 & $ 174 \pm 27 $ & 0.94 & 8.0 & 0.60 & $ 10.82 \pm 0.18 $ & $ 10.52 \pm 0.14 $ 
\enddata
\tablecomments{The slitmask name N10 indicates the objects presented in \citet{newman10}, and the field in which they were observed is given in parentheses. \sigmae\ is the velocity dispersion within one effective radius, calculated using Equation \ref{eq:apcorr2}. The effective radius \R, \Sersic\ index $n$ and axis ratio $q$ are measured in the F160W band. We estimate the observational uncertainty on \R\ to be 10\%. The dynamical masses \Mdyn\ are calculated using Equation \ref{eq:mdyn}.}
\tablenotetext{a}{The structural parameters for these objects are measured in the F814W band instead of F160W.}
\tablenotetext{b}{Objects detected in the X-Ray.}
\end{deluxetable*}


\subsection{SED fitting	}
\label{sec:sed_fit}

We measure stellar masses and other properties by fitting the synthetic stellar population templates from \citet{bruzual03} to the photometric data. We perform the fit using FAST \citep{kriek09}, and adopt the \citet{chabrier03} initial mass function and the \citet{calzetti00} dust extinction law, with attenuations chosen in the range $ 0 < A_V < 3$. We assume an exponentially declining star formation history with timescale $\tau$ and age $t$, and use a logarithmic grid with 10 Myr $ < \tau < 10$ Gyr and 10 Myr $ < t < t_\mathrm{H}$, where $t_\mathrm{H}$ is the age of the universe at the galaxy redshift, fixed to be its spectroscopic value. Because of the well known degeneracy between age and metallicity, we kept the metallicity fixed at the solar value, as appropriate for massive early-type galaxies. For each object, we define the stellar mass and its uncertainty as the mean and standard deviation of the posterior distribution, respectively \citep[see, e.g.,][]{taylor11}. The random uncertainties obtained in this way range from 0.02 to 0.18 dex, with a median of 0.07 dex. However, systematic errors due to, e.g., the choice of IMF and the treatment of AGB stars in the stellar population templates are likely to dominate the uncertainty on stellar masses, particularly at high redshift.

Since we are interested in the relation between the size and the stellar mass of galaxies, we need to ensure that these two quantities are consistently derived. We measure the effective radii via fitting of the surface brightness, assuming a \Sersic\ profile. The flux of the best-fit \Sersic\ model does not necessarily correspond to the flux that one would measure from the same data using a different technique, e.g. adopting a fixed aperture or constructing the curve of growth. This is a particularly relevant issue for our sample, because the SEDs were compiled from different surveys. For this reason we calculated a correction factor in the following way. From the FAST best-fit spectrum we calculate the expected flux $F^{\mathrm{(FAST)}}_{160}$ in the \HST\ filter in which the imaging data have been taken (F160W for most of the objects). We then measure the actual flux $F^{\mathrm{(HST)}}_{160}$ by fitting a \Sersic\ profile to the \HST\ data. Finally, we correct the stellar mass output by FAST: $ \Mstar = \Mstar^\mathrm{(FAST)} \cdot F^{\mathrm{(HST)}}_{160} / F^{\mathrm{(FAST)}}_{160} $. We perform the same correction to the star formation rate, since these are the only parameters that depend on the overall normalization of the observed SED. The correction is generally small, with a mean and standard deviation of $ \langle F^{\mathrm{(HST)}}_{160} / F^{\mathrm{(FAST)}}_{160} \rangle = 0.96 \pm 0.14$. Note that this procedure automatically corrects also for zero point differences among different catalogs, since the corrected stellar masses are normalized to the highly reliable flux calibration of \HST\ data.
The aperture-corrected stellar masses and their uncertainties are reported in Table \ref{tab:sample}. 

\begin{figure}[bp]
\centering
\includegraphics[width=0.45\textwidth]{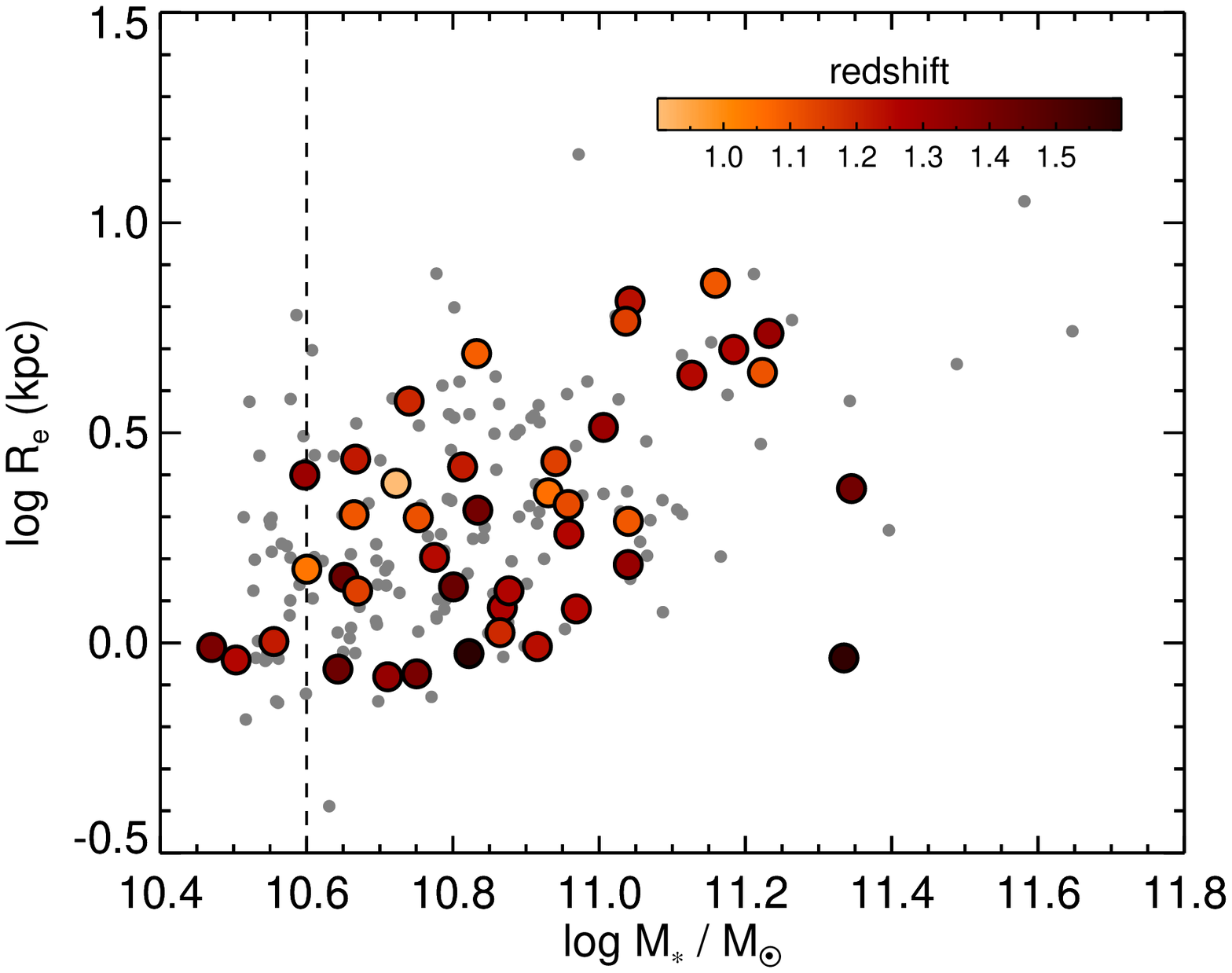}
\caption{Mass-size relation for our spectroscopic sample of quiescent galaxies (large points, color-coded according to their redshift) and for a sample of photometrically selected galaxies with $1<z<1.6$ from \citet[][small gray points]{newman12}.}
\label{fig:candels_comparison}
\end{figure}

Since our spectroscopic sample may be biased toward brighter, more compact objects, we need to check whether completeness effects are important for the subsequent analysis. In Figure \ref{fig:candels_comparison} we compare the stellar masses and effective radii of our sample with those of the quiescent galaxies photometrically selected in the CANDELS fields by \citet{newman12}. Our sample spans the whole range of size for a given stellar mass at all redshifts except for $z>1.5$. Since only two galaxies are in this redshift range, we conclude that our spectroscopic sample is fairly representative of the population of quiescent galaxies with stellar masses above $10^{10.6} \Msun$.

\subsection{Velocity Dispersions}
\label{sec:dispersions}

We derived velocity dispersions by fitting broadened templates to the observed spectra using the Penalized Pixel-Fitting method (pPXF) of \citet{cappellari04}. We used the templates from the \citet{bruzual03} library of synthetic stellar populations, and correct the observed velocity dispersions for instrumental resolution (as measured from unblended sky lines) and template resolution. During the fitting, the pixels near the expected position of the \OII\ emission line were masked, together with the pixels contaminated by strong sky emission. The wavelength region used for the fit is within the range $ 3300 \ \mathrm{\AA} < \lambda < 5500 \ \mathrm{\AA} $ and depends on the rest-frame interval probed by LRIS at each redshift (with the upper limit decreasing with redshift, see Figure \ref{fig:sfit}).

The velocity dispersion $\sigma$ and its uncertainty were calculated as follows. During the template fitting we sum the observed spectrum to a polynomial of degree $m$ to account for template mismatch, and we multiply it by a polynomial of degree $n$ to account for the uncertainty in the relative flux calibration and dust attenuation \citep{cappellari09,bezanson13}. We adopt a grid of polynomial degrees, with $1<m<11$ and $1<n<6$, and calculate the best-fit $\sigma_{mn}$ at each point on the grid. We take as fiducial model the one with $m=8$ and $n=3$, and the corresponding $\sigma$ is our final value of velocity dispersion. Finally, we calculate the uncertainty by summing in quadrature the random error on $\sigma$ output by pPXF in the fiducial model and the standard deviation of $\sigma_{mn}$ after a sigma-clipping on the chi-square distribution to discard poor fits.

We conducted a number of tests to verify that the velocity dispersion measurements are stable and do not depend on the specific assumptions made. The fitting procedure was repeated many times for each object, varying each time one of the parameters. The fraction of pixels discarded due to sky emission does not influence significantly the measured dispersions. We also tested the importance of the template choice. Using the Indo-US library of observed stellar spectra \citep{valdes04} yielded velocity dispersions in good agreement with the ones obtained through the \citet{bruzual03} synthetic spectra of stellar populations, with a median offset of 0.03 dex and a scatter of 0.07 dex. Finally, excluding the calcium H and K lines from the fit does not affect the velocity dispersion measurement in a significant way. We conclude that, for most of the spectra, none of the assumptions involved in the spectral fitting have an influence on the measured dispersions greater than the quoted uncertainties. This is in agreement with the extensive tests performed by \citet{vandesande13} on the spectra of five galaxies at $1.5 < z < 2$.

We discard spectra with a signal-to-noise ratio per resolution element smaller than 8, and we also exclude those galaxies for which the spectral fitting is not stable, i.e., the best-fit parameters change significantly when using higher degree additive and multiplicative polynomials. Our final sample comprises 56 objects (out of 69 examined) on the red sequence, with an average velocity dispersion error $\langle \delta \sigma / \sigma \rangle = 13\%$. This is the largest homogeneous sample of quiescent galaxies at $z>1$ for which reliable velocity dispersions have been measured.

The observed velocity dispersion $\sigma_\mathrm{obs}$ is the luminosity-averaged dispersion within the central region of the galaxy probed by the slit aperture. Since the angular diameter distance and effective radius are different for each object, $\sigma_\mathrm{obs}$ corresponds to different physical regions. To ensure an unbiased comparison, we apply an aperture correction and obtain the velocity dispersion within the effective radius \R. One way to calculate the aperture correction is to adopt the relation between $\sigma_e$ and the velocity dispersion measured within a radius $R$ derived for nearby early-type galaxies by \citet{cappellari06}:
\begin{equation}
\label{eq:apcorr}
	\frac{\sigma_\mathrm{obs} }{ \sigmae } = \left( \frac{ R }{ R_e } \right)^{-0.066} .
\end{equation}
However, at high redshift the effective radius is typically much smaller than the angular size probed by the slit aperture, and the effect of seeing cannot be neglected. The model of \citet{vandesande13}, which takes into account seeing, rectangular aperture, and optimal extraction, is more appropriate for our high-redshift observations. If the seeing is comparable to the slit aperture, as in our case, this model predicts an aperture correction which varies only by 1-2\% with $R/\R$. Therefore we adopt a constant correction factor for all the high-redshift galaxies, taking the average from the \citet{vandesande13} sample:
\begin{equation}
\label{eq:apcorr2}
	\sigmae = 1.05 \; \sigma_\mathrm{obs} .
\end{equation}

The aperture-corrected velocity dispersions are listed in Table \ref{tab:sample}.
For the local comparison sample, we calculate the effective velocity dispersions $\sigma_e$ by applying Equation \ref{eq:apcorr} using $R=1\farcs 5$, corresponding to the radius of the optical fibers used in the SDSS.

\begin{figure*}[tbp]
\centering
\includegraphics[width=0.65\textwidth]{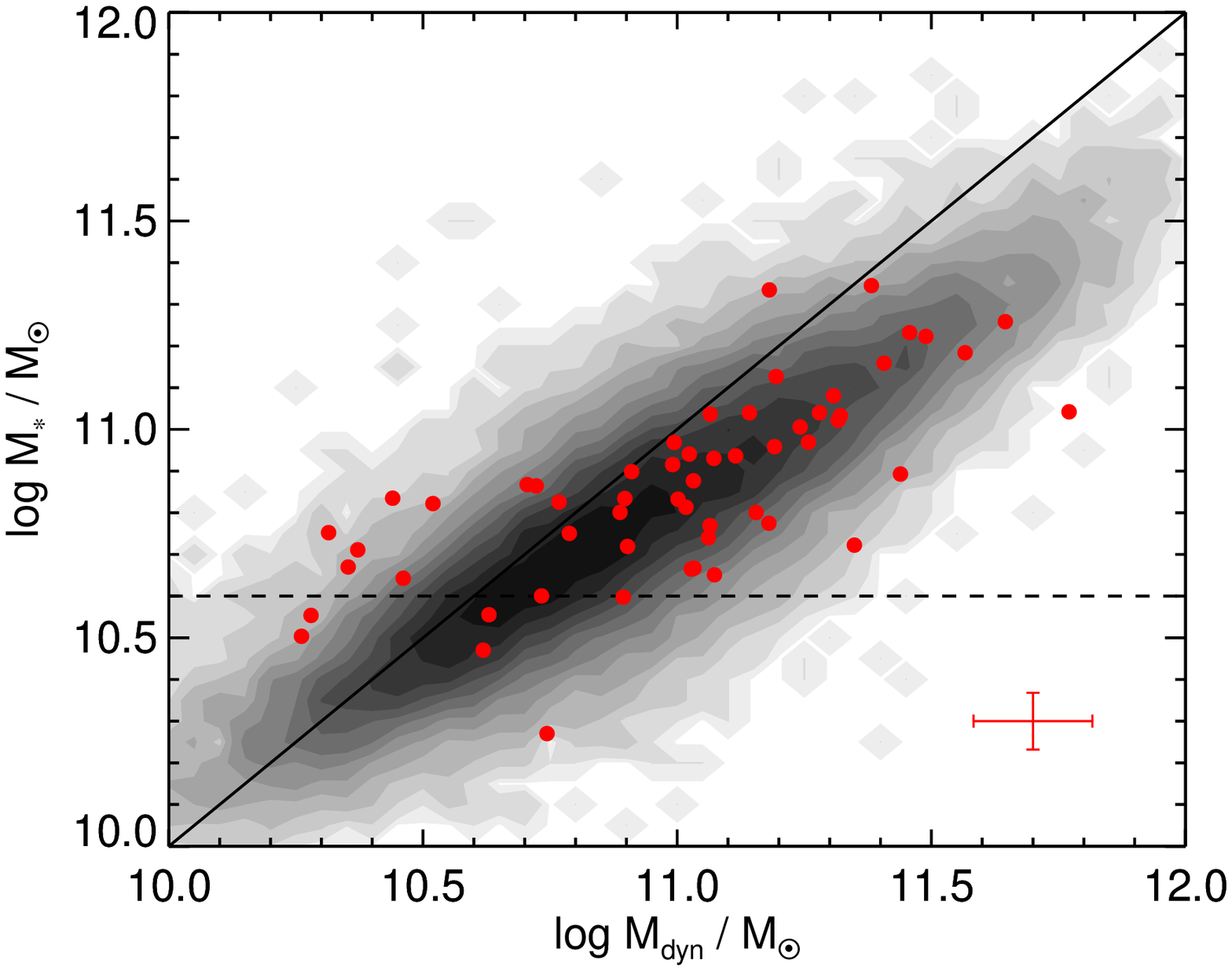}
\caption{Stellar mass versus dynamical mass. Objects from our sample are plotted as red points, and we show the SDSS sample as a grayscale map. The solid line corresponds to equal stellar and dynamical masses, while the dashed line indicates the stellar mass completeness limit. The median error bars on both axis are shown on the bottom right.}
\label{fig:mdyn_mstar}
\end{figure*}

\begin{figure*}[bp]
\centering
\includegraphics[width=0.90\textwidth]{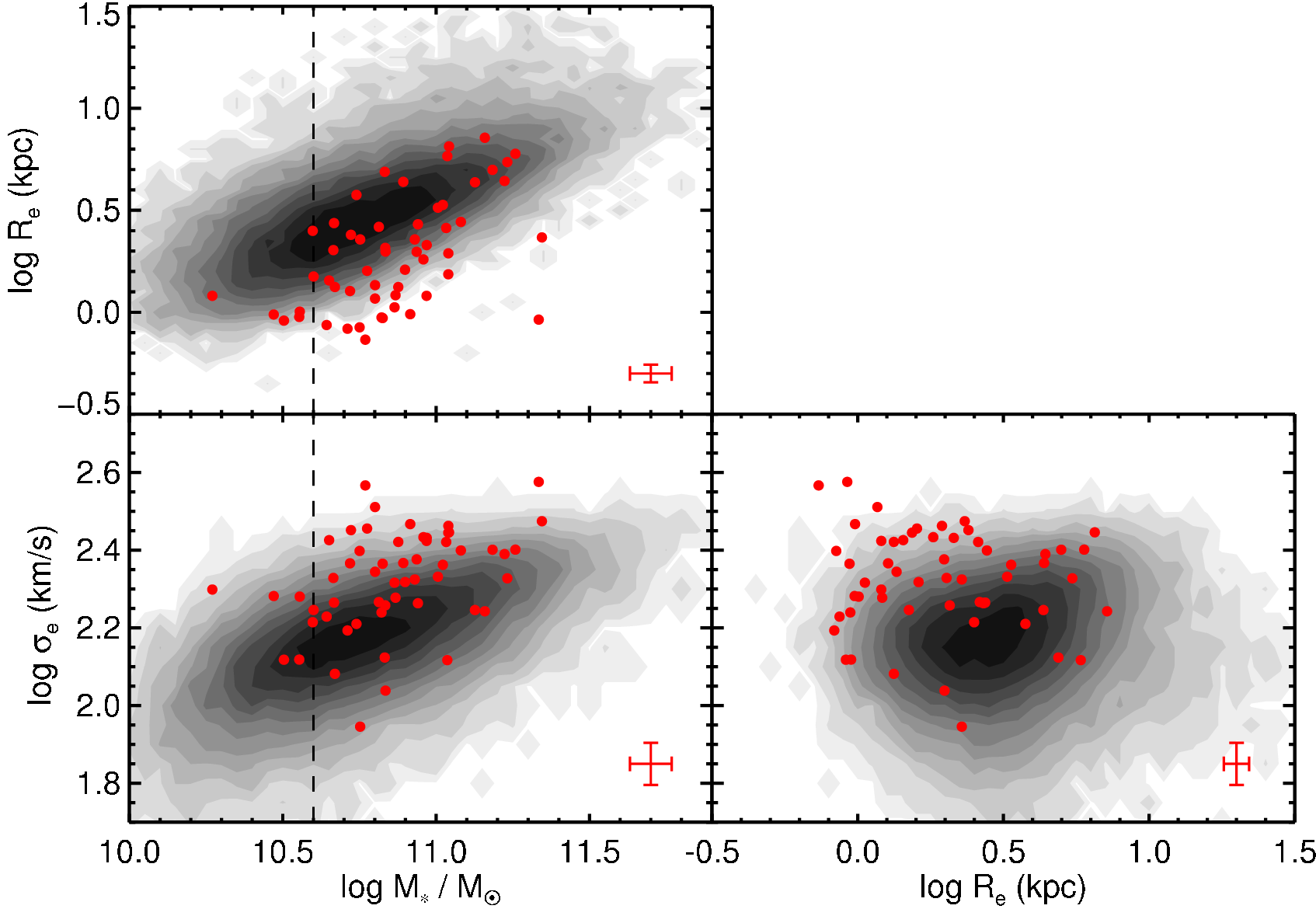}
\caption{\emph{Top:} Effective radius versus stellar mass. \emph{Bottom left:} Velocity dispersion versus stellar mass. \emph{Bottom right:} Velocity dispersion versus effective radius. Symbols as in Figure \ref{fig:mdyn_mstar}.}
\label{fig:sigma_radius}
\end{figure*}


\section{Dynamical Masses}
\label{sec:dynmasses}

We now turn to determining dynamical masses for our sample. From a simple virial argument, it is possible to relate the dynamical mass of a galaxy to its velocity dispersion \sigmae\ and effective radius \R: $G\Mdyn = \beta \sigmae^2 \R$, where the virial factor $\beta$ depends on the galaxy structure. \citet{cappellari06} showed that a constant $\beta = 5$ is a good approximation for elliptical galaxies. We then define the dynamical mass as
\begin{equation}
\label{eq:mdyn}
	\Mdyn = \frac{5 \sigmae^2 \R}{G} .
\end{equation}

Via this procedure we determined dynamical masses for our sample and list these in Table \ref{tab:sample}. We also calculate dynamical masses for the local SDSS sample using Equation \ref{eq:mdyn}. 

We note that the use of \Sersic\ profiles to describe the surface photometry implies that galaxies with different indices $n$ will naturally have different structures and therefore different virial factors. We explore this topic further in Appendix \ref{appendix:sersicn}.

\subsection{The Stellar Mass-Dynamical Mass Relation}

In Figure \ref{fig:mdyn_mstar} we compare the stellar masses \Mstar\ and dynamical masses \Mdyn\ for our sample of quiescent galaxies (red points) and for the SDSS sample (grayscale map). The relation $\Mdyn = \Mstar$ is shown by the black line. The region above this line indicates the unphysical situation where the stellar mass exceeds the dynamical mass. 

It is clear that the high-redshift quiescent galaxies occupy the same region as the local population, except for a group of outliers at low masses, near our completeness threshold. Overall, the correlation between stellar and dynamical mass in Figure \ref{fig:mdyn_mstar} is very good at both high and low redshift. In Figure \ref{fig:sigma_radius} we develop the case further by showing the relation between stellar mass, size, and velocity dispersion for quiescent galaxies. Here, the high-redshift population shows a significant offset from the SDSS sample: even though there is some overlap with the local population, the systematic shift is clear. At fixed stellar mass, their effective radii are significantly smaller, and their velocity dispersions larger, compared to the local population. To quantify this offset approximately, we perform a linear fit to the SDSS data in the range $ 10.5 < \log \Mstar/\Msun < 11.5 $, then fit a line with the same slope to the high-redshift data points. The offset obtained in this way is $-0.25 \pm 0.03$ dex in size and $+0.12 \pm 0.02$ dex in velocity dispersion. According to Equation \ref{eq:mdyn}, an offset in size and velocity dispersion will produce a shift in the stellar-dynamical mass relation equal to $2 \Delta \log \sigmae + \Delta \log \R $. The measured offsets, then, cancel each other almost exactly, leaving the ratio of stellar to dynamical mass unchanged, as seen in Figure \ref{fig:mdyn_mstar}. This fact is noteworthy for two reasons. 

Firstly, from an observational point of view, it confirms the validity of our measurements. Since sizes, velocity dispersions and stellar masses are measured from, respectively, \HST\ imaging, Keck spectroscopy and broad-band photometry, these three key observables are effectively independent (since the stellar mass aperture corrections, derived from the imaging data, are small; see Section \ref{sec:sed_fit}). If any one of these were to be biased because of some observational effect, then a fine-tuned bias in the other two quantities would be required to produce the agreement seen in Figure \ref{fig:mdyn_mstar}. Thus, importantly, the relatively large velocity dispersions measured are a confirmation of the small sizes of high-redshift quiescent galaxies.

Secondly, the fact that the offsets in size and velocity dispersions do not produce an offset in dynamical mass has important implications for the evolution of early-type galaxies between $z\sim1.3$ and today. We explore this further in the next subsection.

\subsection{The Redshift Evolution of the Mass Ratio}

\begin{figure*}[t]
\centering
\includegraphics[width=0.60\textwidth]{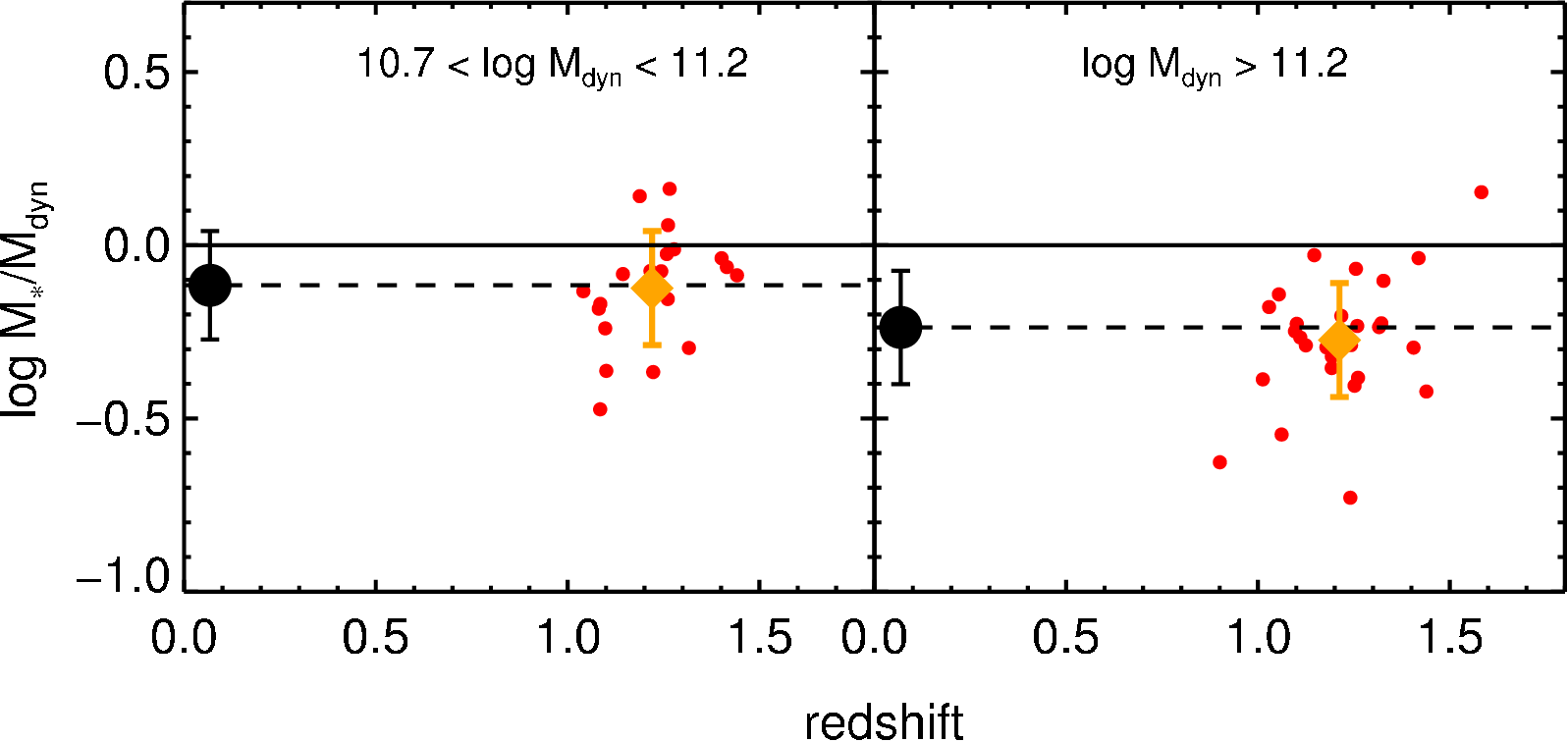}
\caption{Redshift evolution of the stellar-to-dynamical mass ratio, for intermediate (left) and high mass (right) galaxies. The black points represent the SDSS sample, and the red points are galaxies from our sample, for which the average values and standard deviations are shown in orange.}
\label{fig:massratio}
\end{figure*}

As seen in Figure \ref{fig:mdyn_mstar}, the distribution of the mass ratios \Mstar/\Mdyn\ is very similar for the high and low-redshift populations. Both samples have similar average ratios: $ \langle \log \left( \Mstar / \Mdyn \right) \rangle = -0.13 $ for our sample and $-0.12$ for the SDSS population. The scatter is slightly larger at $z>1$, with the standard deviation being 0.25 dex at high redshift and 0.18 dex at $z\sim0$. In Figure \ref{fig:massratio} we consider the redshift evolution of the ratio \Mstar/\Mdyn\ into two mass bins. The mean mass ratio for the high-redshift sample agrees with the local value in both bins. The standard deviation is also approximately unchanged from $z\sim1.3$ to $z\sim0$. Although \citet{vandesande13} found evidence for a slight evolution of the mass ratio at $z > 1.5$,  within our larger sample we find no significant evolution in the relation between stellar and dynamical mass for quiescent galaxies over $0 < z < 1.6$.

Since the dynamical masses are derived independently of the synthetic stellar populations, the absence of a systematic offset in the two distributions suggests that the stellar masses at $z\sim1.3$ are reliable. However, we cannot exclude some evolution in the intrinsic mass ratio together with a bias in the stellar masses, e.g. one caused by evolution in the initial stellar mass function (IMF) for the recently quenched galaxies,  that conspire to produce this result. Nonetheless, our data are consistent with the simplest possible scenario, in which both IMF and dark matter fraction are unchanging over $ 0 < z < 1.6$. Of course, since galaxies evolve in mass and size with time, the fact that the stellar-dynamical mass relation is constant with redshift does not necessarily imply no evolution in the ratio \Mstar/\Mdyn\ {\it for individual objects}. We will explore this point further in Section \ref{sec:size_evo}.

Finally, we use our data to test the scenario proposed by \citet{peraltadearriba13}, in which compact galaxies present dynamical masses significantly smaller than their stellar masses, an unphysical situation which they attributed to a strong non-homology in galaxy structure. From Figure \ref{fig:mdyn_mstar} we can see that the majority of high-redshift galaxies in our sample have $\Mdyn > \Mstar$, and the few exceptions lie near the completeness limit. We therefore rule out a discrepancy between dynamical and stellar mass measurements. However, we do find a clear correlation between compactness (more precisely, velocity dispersion) and mass ratio, which we will further explore in Section \ref{sec:massratio_evo}.

\subsection{Testing Inferred Velocity Dispersions}

Spectroscopic measurements of the stellar velocity dispersion for high-redshift galaxies require very long integrations and so far have been performed on a small number of objects. In fact, the present sample is the largest at redshifts $z > 1$. A more economic approach is to estimate the velocity dispersion from photometric data using a local calibration \citep{bezanson11}. Although this produces inferred velocity dispersions for large samples, it relies on the assumption that the local calibration is valid at all redshifts. Our spectroscopic sample presents a unique opportunity for testing this assumption.

Following \citet{bezanson11}, we use Equation \ref{eq:mdyn} to define the \emph{inferred} velocity dispersion as
\begin{equation}
\label{eq:sigma_inf}
	\sigmainf = \sqrt{ \frac{ G }{ 5 \; \R } \; 0.15 \; \Mstar^{1.09} } ,
\end{equation}
where $\Mdyn=0.15 \; \Mstar^{1.09}$ is the result of a linear fit to the SDSS galaxies in the mass range $10^{10.5}$-$10^{11.5}$\Msun. This equation differs from the one given by \citet{bezanson11} because we do not include the dependence of the virial factor on the \Sersic\ index, which we discuss in Appendix \ref{appendix:sersicn}.

In Figure \ref{fig:sigma_inf} we plot inferred versus spectroscopic velocity dispersions for our sample and for the SDSS local population. There is good agreement at all values of velocity dispersion, including for the very large ones, which are poorly sampled in the local distribution. The scatter is 0.13 dex and is slightly larger than the one found in the $z\sim0$ population, which is 0.10 dex. This difference most likely arises as a result of greater observational uncertainties at high redshift. We conclude that the local, empirical calibration for determining inferred dispersions holds reasonably well for galaxies at $1 < z < 1.6$. This can be explained physically as a  consequence of the observed constancy of the relation between stellar and dynamical masses (Figure \ref{fig:mdyn_mstar}). However, the scatter of 0.13 dex, or about 35\%, is much larger than the  13\% typical uncertainty on the spectroscopic dispersions. This clearly limits the precision of the inferred dispersions rendering this method less useful except for statistical studies of large populations. 

\begin{figure}[tbp]
\centering
\includegraphics[width=0.45\textwidth]{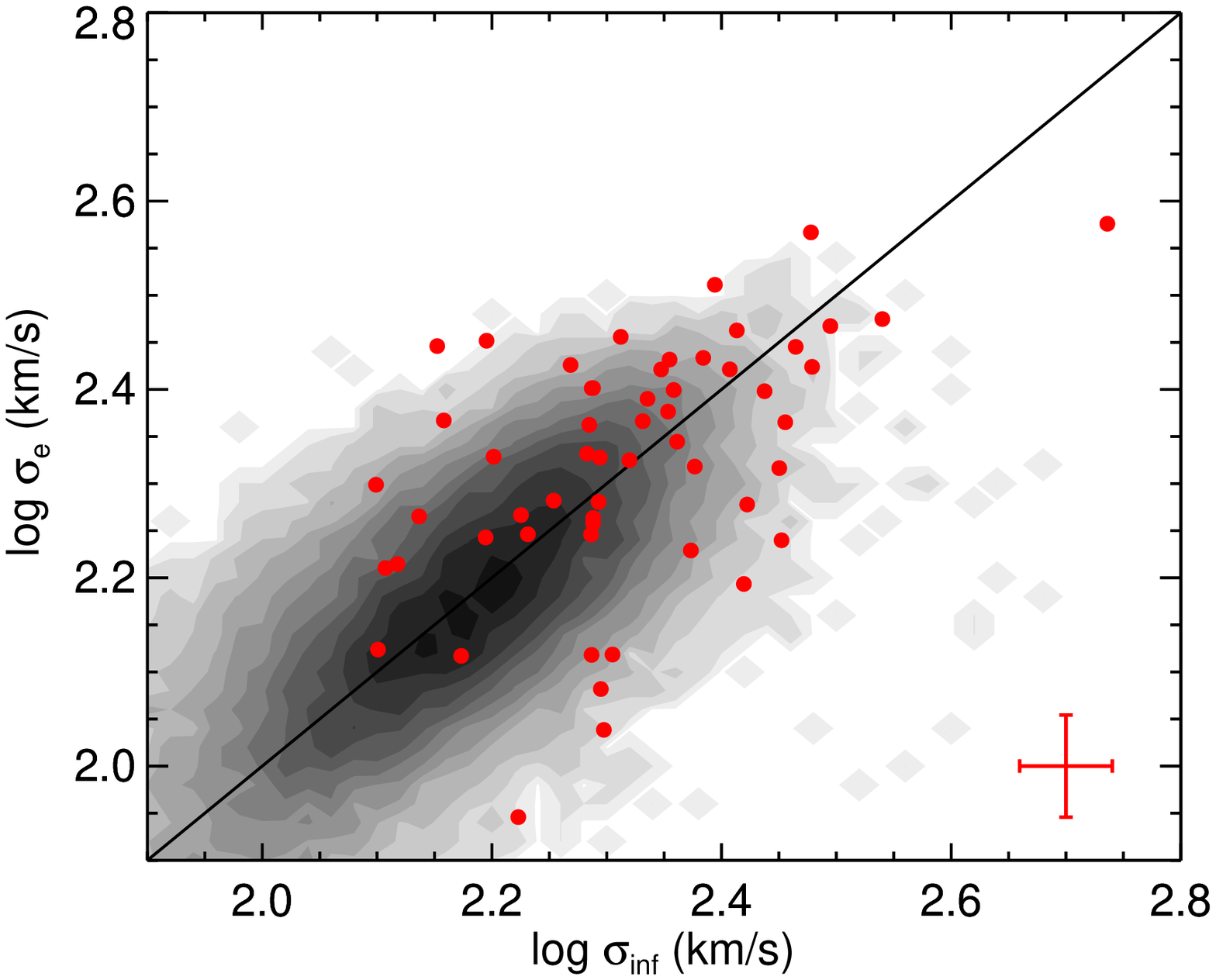}
\caption{Spectroscopically observed versus photometrically inferred velocity dispersions. Red points represent high-redshift quiescent galaxies, and the grayscale map is the distribution of the SDSS sample. The black line is the 1:1 relation. The median error bars are shown on the bottom right corner.}
\label{fig:sigma_inf}
\end{figure}


\section{The size growth of quiescent galaxies}
\label{sec:size_evo}

In the previous section we presented a clear difference between the sizes and velocity dispersions of local and high-redshift galaxies, yet noted the remarkable constancy of the overall stellar-to-dynamical mass ratio. We will now explore in more detail the size evolution that high-redshift galaxies must undergo in order to match the observed properties of the $z\sim0$ population.

\subsection{The Progenitor Bias}

There are two important effects that we need to take into account when modeling the evolution of quiescent galaxies. Firstly, even though these objects form very little stars, they can change their stellar mass and other properties through galaxy merging. Secondly, newly quiescent galaxies are continually being added to the red sequence as blue galaxies shut down their star formation and turn into red, quiescent objects. This quenching process is responsible for a population of local early-type galaxies that were not quiescent at $z>1$. This mismatch in identification between low- and high-redshift galaxy populations is comprehensively called progenitor bias.

The effect of galaxy merging differs according to the ratio of the masses involved:
\begin{itemize}
\item \emph{Major merging}, i.e. merging between two galaxies of similar mass, reduces the number density and also has a large effect on the mass and size of the galaxies. Theoretical arguments and numerical simulations \citep[e.g.,][]{hernquist93,naab09,hilz13} predict that in major mergers the size and the stellar mass of individual galaxies grow at the same rate: $\R \propto \Mstar$.
\item \emph{Minor merging}, on the other hand, does not have a large effect on the stellar mass of a galaxy, but can alter its size. In this case the theoretical expectation is a size growth steeper than that caused by major merging: $\R \propto \Mstar^\alpha$, with $1<\alpha<2.5$.
\end{itemize}
The combination of these processes makes it very difficult to identify, for a given high-redshift galaxy, its potential descendants in a $z\sim0$ population.

Moreover, the quenching of star-forming galaxies introduces the complementary issue of finding, in a low redshift population, those galaxies whose progenitors were already quiescent at $z>1$. In fact, it has been suggested that the observed discrepancy between the sizes of local and high-redshift quiescent galaxies could be fully explained by the progenitor bias \citep[e.g.,][]{carollo13,poggianti13numberdensity}. In this scenario, little physical size growth of individual galaxies is required, since the larger average radius observed at $z\sim0$ can primarily be due to the contribution of recently formed quiescent galaxies. 

\subsection{Evolution at Fixed Velocity Dispersion}
\label{subsec:fixedsigma}

Comparing the sizes of low and high-redshift galaxies at fixed stellar mass is not particularly helpful in understanding the physical evolution of individual galaxies, since stellar masses can significantly increase after, e.g., a major merger. Taking advantage of our unique spectroscopic dataset, we therefore compare galaxy sizes \emph{at fixed velocity dispersion}. There are several reasons why the stellar velocity dispersion is thought to remain relatively constant with cosmic time. From an observational point of view, \citet{bezanson12} showed that the number density of galaxies with large (inferred) velocity dispersion changed very little since $z\sim1.5$. Also, numerical simulations show that the central velocity dispersion is weakly affected by minor or major mergers, and changes by only 10\% from $z\sim2$ to $z\sim0$ \citep[e.g.,][]{hopkins09scalingrel}.

The lower panels of Figure \ref{fig:sigma_radius} show that high-redshift galaxies have larger velocity dispersions and smaller radii compared to the typical SDSS values. In particular, at fixed stellar mass the velocity dispersions at high redshift are higher, as we discussed in the previous section. Assuming a constant \sigmae, high-redshift galaxies are constrained to evolve along horizontal tracks, therefore they are not able to evenly populate the distribution of velocity dispersion observed locally. The only way to reproduce the local distribution would be to assume that all the newly quenched galaxies lie in the lower \sigmae\ region of the figure. Thus it follows that, at $z\sim0$, older galaxies have large velocity dispersions.  \citet{graves09_I} \citep[and others, e.g.,][]{thomas05} studied the stellar populations of SDSS early-type galaxies and found a convincing correlation between velocity dispersion and age. In particular, they concluded that all galaxies with $\log \sigmae > 2.35$ (aperture-corrected to our system) are older than 10 Gyr, corresponding to a formation epoch earlier than $z = 1.6$. A self-consistent picture emerges: at high redshift we observe quiescent galaxies that locally have old stellar populations and large velocity dispersions. This agrees with the simple analytical model of \citet{vanderwel09}, whereby all early-type galaxies with $\log \sigmae = 2.40$ formed at $z\sim1.5$.

\begin{figure}[tbp]
\centering
\includegraphics[width=0.45\textwidth]{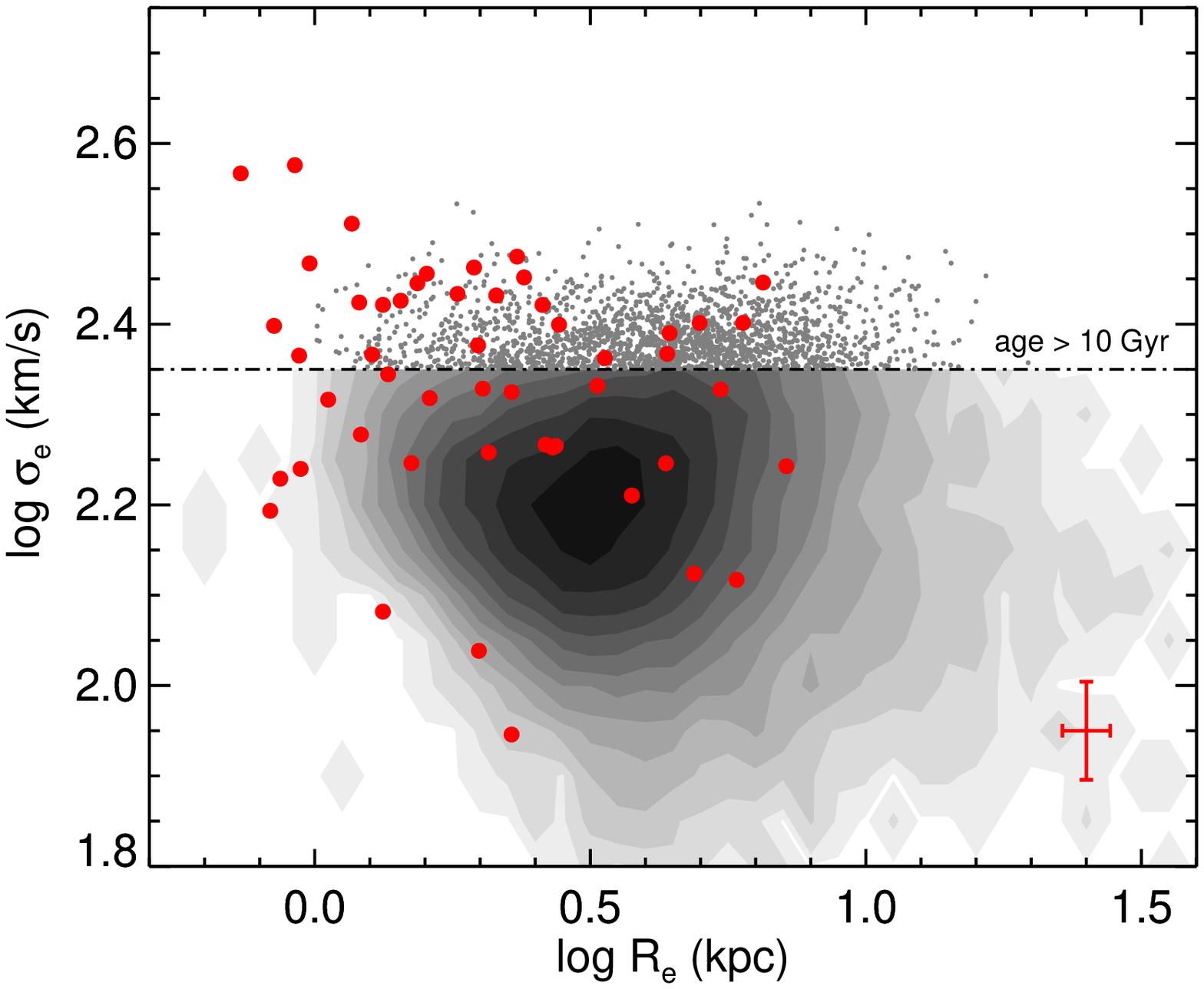}
\caption{Velocity dispersion versus effective radius, for galaxies with $\log\Mstar/\Msun > 10.6$. High-redshift galaxies are shown as red points, and the SDSS sample as a grayscale map. The dot-dashed line marks the velocity dispersion above which, in the local universe, galaxies are older than 10 Gyr \citep{graves09_I}. In this region, SDSS galaxies are plotted individually as gray small points. The median error bars for the high-redshift sample are shown on the bottom right corner.}
\label{fig:sigma_radius_age}
\end{figure}

In Figure \ref{fig:sigma_radius_age} we re-examine the \R-\sigmae\ plane for low and high-redshift galaxies, this time plotting only the galaxies above our completeness limit, $\log \Mstar/\Msun > 10.6$. The horizontal dot-dashed line represents the $\log \sigmae = 2.35$ threshold from \citet{graves09_I}. In the region above this threshold we plot individual SDSS galaxies to better facilitate the comparison with the high-redshift sample. Since all the SDSS points above the line have very old stellar populations, we conclude that it is reasonable to connect the two distributions. The difference in size between the red and gray points is more than a factor of 2 (the mean offset is significant, viz $0.33 \pm 0.05$ dex), and cannot be accounted for by recently quenched galaxies of younger ages. High-redshift quiescent galaxies must \emph{physically grow} in size in order to match the local distribution.

There is further independent evidence that progenitor bias is insufficient to explain the observed size evolution. Within the SDSS sample, \citet{graves09_II} and \citet{vanderwel09} independently found no correlation between size and age for quiescent galaxies at fixed velocity dispersion. In other words, considering Figure \ref{fig:sigma_radius_age}, all the $z\sim0$ galaxies along a horizontal line have similar ages. Not only does this confirm that larger radii do not correspond to more recently-quenched galaxies, but it also extends the test to lower velocity dispersions. Since the red points preferentially occupy the portion of the figure corresponding to smaller radii also at $\log \sigmae < 2.35$, then physical growth is essential as otherwise size and age would correlate in the SDSS sample. It is worth noting that the lack of a size-age relation holds at fixed velocity dispersion, but not at fixed stellar mass \citep{vanderwel09}. 

Finally, considering the two left panels of Figure \ref{fig:sigma_radius}, if the velocity dispersions remain constant and the sizes increase from $z>1$ to $z\sim0$, then the stellar masses must likewise increase to reproduce the local distribution.

To summarize, we have constructed a simple model for the evolution of quiescent galaxies over $0<z<1.6$ based on the following assumptions:
\begin{enumerate}
\item The velocity dispersions of individual galaxies do not change with cosmic time.
\item In the local universe, galaxies with larger velocity dispersions are older \citep{graves09_I}, while at fixed velocity dispersion there is no correlation between size and age \citep{graves09_II}.
\end{enumerate}
With these assumption, observations of quiescent galaxies at low and high redshifts can be reconciled only if the sizes of individual quiescent galaxies physically grow with cosmic time. 
In the following, the implications of this simple model are considered.

\subsubsection{The Evolution of Dynamical and Stellar Masses}
\label{sec:massratio_evo}

\begin{figure}[bp]
\centering
\includegraphics[width=0.45\textwidth]{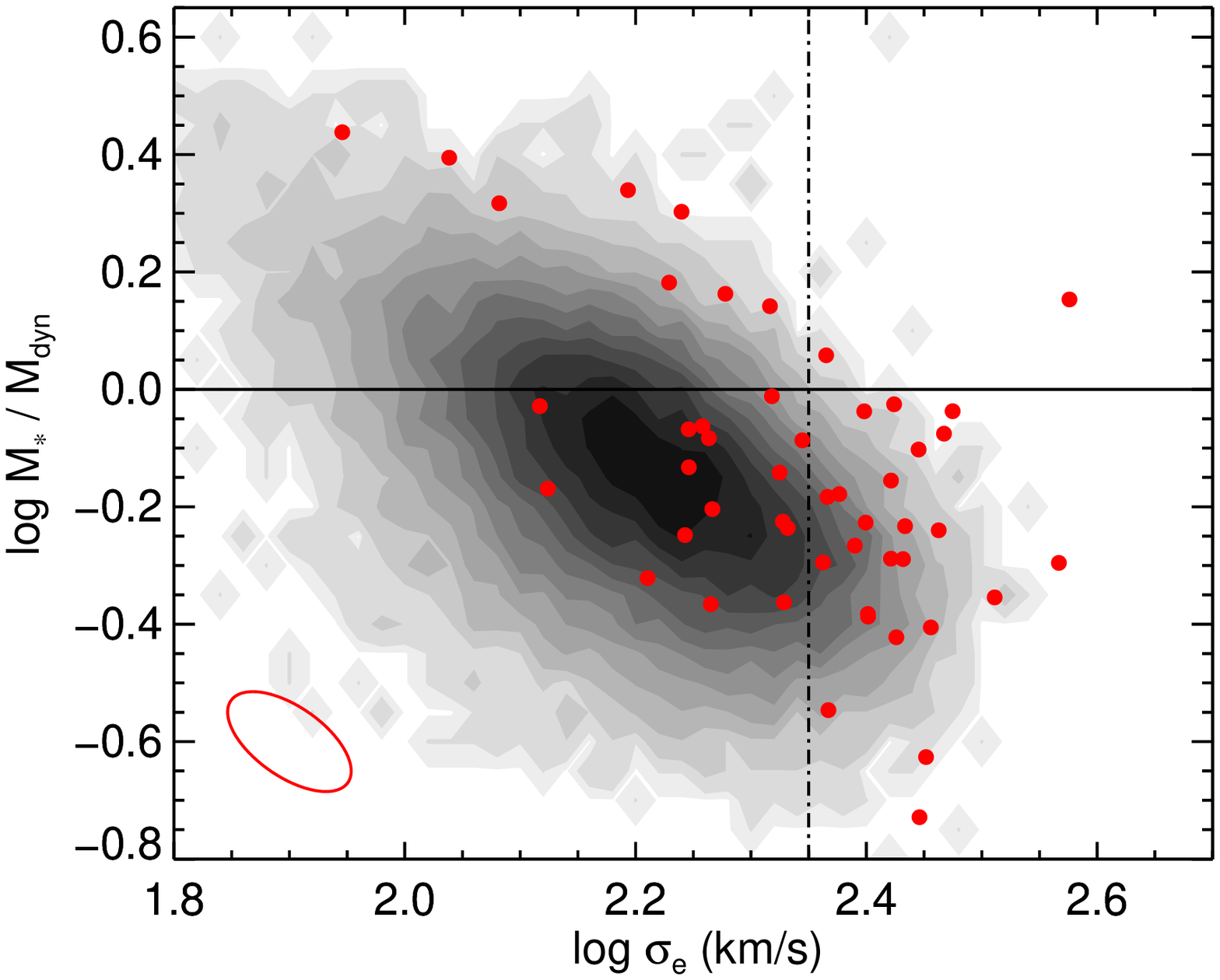}
\caption{Stellar-to-dynamical mass ratio versus velocity dispersion, for galaxies with $\log\Mstar/\Msun > 10.6$. The grayscale map represents the SDSS sample, and the red points are high-redshift galaxies from our sample. The dot-dashed line indicates the threshold velocity dispersion above which galaxies are older than 10 Gyr. The median error ellipse for the high-redshift sample is shown in the bottom left corner.}
\label{fig:mratio_sigma}
\end{figure}

A very interesting quantity which can be studied using our new data is the stellar-to-dynamical mass ratio. In the local universe this mass ratio shows an inverse correlation with velocity dispersion \citep[e.g.,][]{taylor10}. Figure \ref{fig:mratio_sigma} shows this relation for both local and high-redshift samples with the restriction to galaxies more massive than our completeness limit, $\log \Mstar/\Msun > 10.6$. It is important to note that the two axes are not independent, since the dynamical mass depends on velocity dispersion. Correlated errors lead to a preferred direction for the scatter of the data points as shown in the median error ellipse, calculated assuming normally distributed errors on stellar mass, radius, and velocity dispersion.

\begin{figure*}[tbp]
\centering
\includegraphics[width=0.45\textwidth]{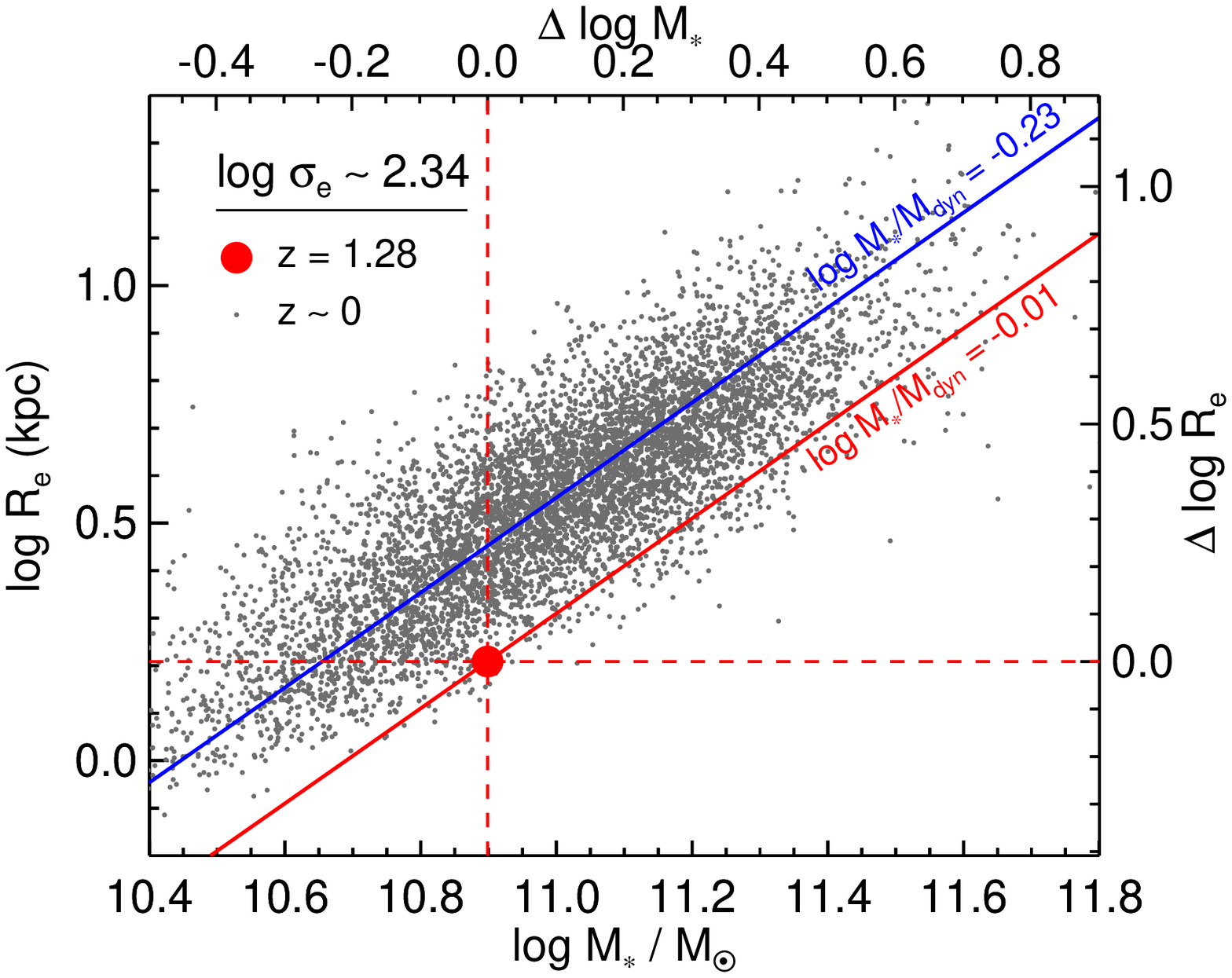}
\includegraphics[width=0.45\textwidth]{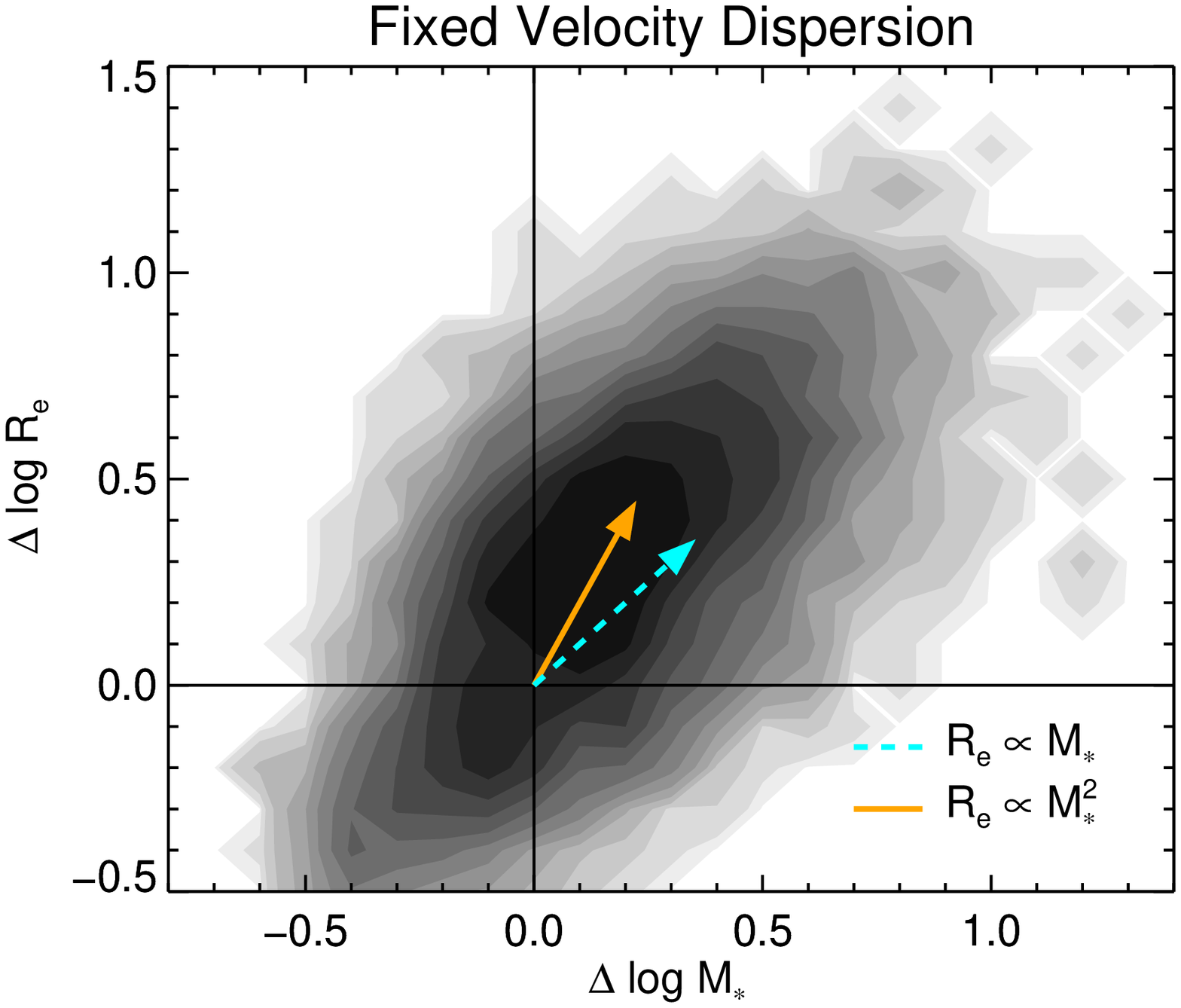}
\caption{Matching at fixed velocity dispersion. \emph{Left:} Mass-size relation. The red point is a $z = 1.28$ object in our sample with $\log \sigmae = 2.34$, shown as example. The gray points are all the SDSS galaxies with same velocity dispersion, within 0.025 dex. The top and right axis show the logarithmic offset from the high-redshift point in stellar mass and effective radius, respectively. The diagonal lines represent fixed stellar-to-dynamical mass ratios. The blue line corresponds to the median mass ratio of the SDSS sample, and the red line corresponds to the mass ratio of the high-redshift object. \emph{Right:} Comparison of stellar mass and effective radius at fixed velocity dispersion for the whole sample. This plot is constructed by stacking the matched low-redshift population offsets (like the one shown on the left panel) for all the LRIS objects with $\log \Mstar/\Msun > 10.6$. The dashed cyan and solid orange arrows represent two example of size growth: $\R \propto \Mstar$ and $\R \propto \Mstar^2$, respectively.}
\label{fig:fixed_sigma}
\end{figure*}

Clearly galaxies with larger \sigmae\ tend to have smaller \Mstar/\Mdyn\ ratios. Also, the high-redshift trend is offset from the local relation toward larger \sigmae\ and larger mass ratios. Correlated errors can only be partly responsible for the trend seen among the red points in Figure \ref{fig:mratio_sigma}, as the sequence spans 0.6 dex in \sigmae\ and 0.8 dex in mass ratio, a range much larger than the observational uncertainties. Furthermore, as the error ellipse is orthogonal to the shift of the high-redshift sequence with respect to the local one, we conclude that the observed offset is real.

Figure \ref{fig:mratio_sigma} also marks the threshold velocity dispersion above which local galaxies have stellar populations older than 10 Gyr. To the right of the dot-dashed line, high-redshift galaxies are required to evolve until they match the SDSS distribution. Our high-redshift quiescent galaxies have slightly larger $\Mstar/\Mdyn$ ratios: at $\log \sigmae > 2.35$ the two samples are offset by 0.05 dex. This signal is not as clear as in the velocity dispersion - size distribution, likely because of the effect of correlated errors. We expect, nevertheless, a mild evolution of the mass ratio of individual galaxies, since effective radius and stellar mass, as we showed, evolve with redshift.

Galaxies with velocity dispersion smaller than the threshold value are in general less constrained by our observations. Recently quenched objects could occupy preferentially the lower region of the figure, and in this scenario high-redshift galaxies would not be required to evolve and match the SDSS population. However, if their sizes and masses are evolving, then their mass ratio will change with time, except in the particular case of mass and size growing at the same rate, since $\Mstar/\Mdyn \propto \Mstar/\R$, at fixed velocity dispersion. This would correspond to evolution driven by major merging that does not affect the mass ratio since both masses change by the same amount. The evolutionary tracks would then be parallel to the one-to-one relation on the \Mstar\ versus \Mdyn\ relation in Figure \ref{fig:mdyn_mstar}.

In order to constrain the evolution of high-redshift galaxies, once again we make use of the results from studies of the local universe. According to \citet{graves10_III}, at fixed velocity dispersion older galaxies have \emph{lower} \Mstar/\Mdyn\ ratios. This means that the high-redshift points in Figure \ref{fig:mratio_sigma} need to evolve toward lower mass ratios until they match the SDSS distribution, and potentially even further, in order to populate the bottom of the plot. Therefore we rule out a scenario in which stellar and dynamical mass grow at the same rate and the mass ratio of individual galaxies does not change.

Finally, we emphasize that the evolution in the mass ratio of individual galaxies is not in contradiction with the redshift-independent sequence in the stellar-dynamical mass plane. A given position on the sequence can be populated both by a galaxy at $z>1$ and a galaxy at $z\sim0$: they will have same stellar mass, dynamical mass, and mass ratio, but they will differ in velocity dispersion and, therefore, effective radius. According to our model, the high-redshift galaxy will then evolve at fixed velocity dispersion, increasing its size and stellar mass, but remain on the \Mstar-\Mdyn\ sequence. At $z\sim0$ it will occupy a different region of the sequence and will have a smaller \Mstar/\Mdyn\ ratio. This scenario is in qualitative agreement with the prediction of numerical simulations of minor merging \citep[e.g.,][]{hopkins09scalingrel,hilz13}.

\begin{figure*}[tbp]
\centering
\includegraphics[width=0.45\textwidth]{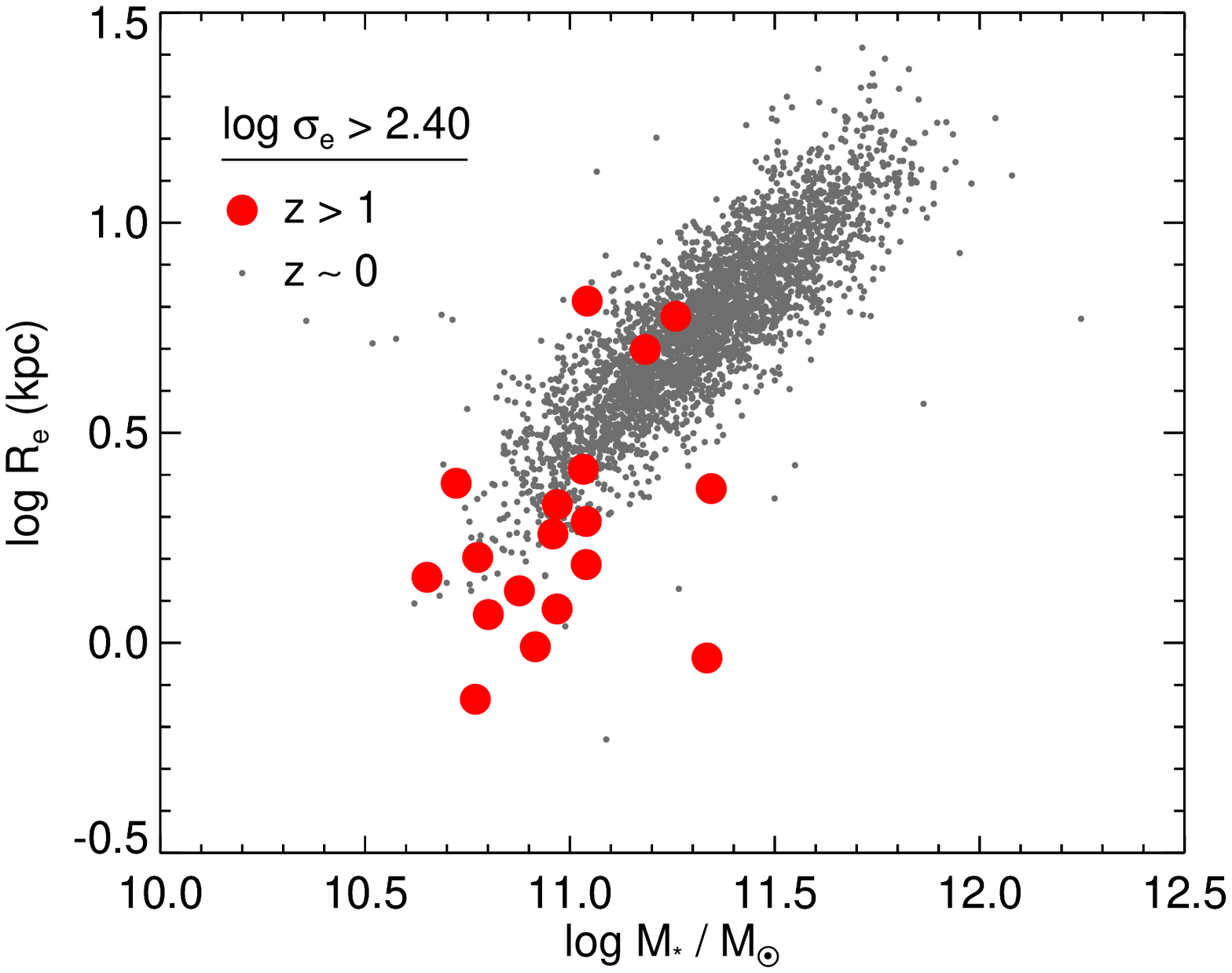}
\includegraphics[width=0.45\textwidth]{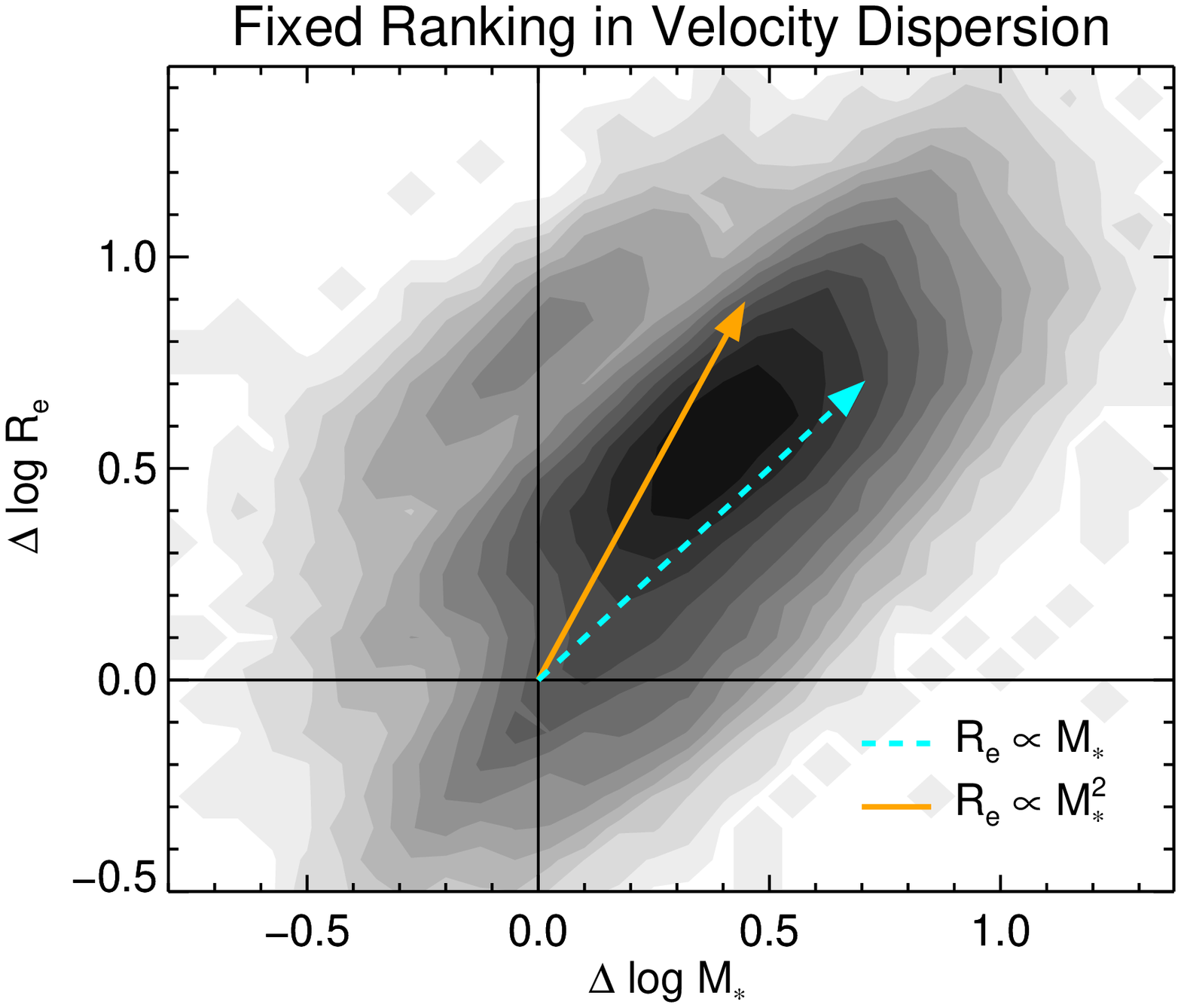}
\caption{Matching at fixed velocity dispersion ranking. \emph{Left:} Stellar mass - size plane for all the high-dispersion galaxies, defined by $\log \sigmae > 2.40$, at high (red points) and low (gray points) redshift. \emph{Right:} Inferred evolution in size and stellar mass of high-dispersion galaxies between $z\sim1.3$ and $z\sim0$. Each point represents the logarithmic offset in radius and mass of a galaxy in the SDSS sample compared to a high-redshift galaxy, both with $\log \sigmae > 2.40$. The dashed cyan and solid orange arrows represent two examples of size growth: $\R \propto \Mstar$ and $\R \propto \Mstar^2$, respectively. The length of the arrows is twice the length of the arrows in Figure \ref{fig:fixed_sigma}.}
\label{fig:fixed_ranking}
\end{figure*}

\subsubsection{Inferring the Size Growth}

We can now take full advantage of our high-quality spectroscopic data set and derive quantitatively the physical evolution in size and stellar mass of quiescent galaxies over $0<z<1.6$. Since we are assuming that the velocity dispersions of individual galaxies do not change with time, a natural choice is to compare the physical properties of each high-redshift galaxy of dispersion $\sigma_0$ with a subsample of the local population selected to have $\sigma_0 - h < \sigma < \sigma_0 + h$, where $h$ is a small bin size (we take $h = 0.025$ dex). An example is shown in the left panel of Figure \ref{fig:fixed_sigma}: the red point is a high-redshift galaxy and the gray points represent the SDSS sample selected to have a similar velocity dispersion. Here we are assuming that the red point will physically evolve to become any one of the gray points at $z\sim0$. 
This allows us to determine the mean growth in size. Note that some of the $z\sim0$ sample will be composed of galaxies that were quenched only recently, and therefore have no quiescent progenitors at $z>1$. However, this should not bias significantly the results of our analysis since, as we discussed previously, at fixed velocity dispersion there is no correlation between size and age \citep{graves09_II}, and therefore we can assume that young and old galaxies are evenly distributed among the SDSS points in the figure.

The $z\sim0$ population forms a tight sequence in the mass-size plane which arises because of the relation between stellar and dynamical mass discussed previously. At fixed velocity dispersion we have:
\begin{equation}
\label{eq:mass-size}
	\log \Mstar = \log \Mstar/\Mdyn + \log 5/G + 2 \log \sigmae + \log \R .
\end{equation} 
Assuming a constant stellar-to-dynamical mass ratio (within a limited range in stellar mass), we obtain a linear relation between stellar mass and effective radius. The linear relation that fits the median of the SDSS sample is shown in blue in Figure \ref{fig:fixed_sigma}. The median mass ratio for this sample is $\log \Mstar/\Mdyn = -0.23$. The linear relation corresponding to the high-redshift galaxy that we took as example (shown in red) is offset by more than 0.2 dex from the local relation, despite the fact that this galaxy has, by construction, the same velocity dispersion as the $z\sim0$ sample. As we discuss below this offset gives us a method to determine the growth in size and mass. The shift arises via the difference in the mass ratio: in Section \ref{sec:massratio_evo} we demonstrated that, at a given \sigmae, high-redshift galaxies have a higher mass ratio than local galaxies. For example, in this particular case the high-redshift galaxy has $\log \Mstar/\Mdyn = -0.01$. We also showed that the stellar and dynamical masses of high-redshift galaxies change with time, and their mass ratio decreases. Looking at the left panel in Figure \ref{fig:fixed_sigma}, this means that the red point must evolve onto the blue line.

For each object in the local matched sample the offset in size, $\Delta \log \R $, and stellar mass, $\Delta \log \Mstar $, from the high-redshift galaxy is calculated (shown in the top and right axis in the left panel of Figure \ref{fig:fixed_sigma}). This exercise is repeated for all the galaxies in the high-redshift sample with $\log \Mstar/\Msun > 10.6$, and the distribution of mass and size offsets is summed as in the right panel of Figure \ref{fig:fixed_sigma}. Since the distribution of \sigmae\ in the SDSS sample falls steeply with increasing values, high-redshift galaxies with lower velocity dispersion have generally a larger number of $z\sim0$ matching objects. In order to ensure even weighting, for each high-redshift galaxy we randomly draw from the matched $z\sim0$ sample a fixed number of objects (250). As a small number of high-redshift galaxies have a velocity dispersion higher than any local galaxy (see Figure \ref{fig:sigma_radius_age}), we temporarily exclude them from the analysis.

The right panel of Figure \ref{fig:fixed_sigma} can be interpreted as the probability distribution that a galaxy at $1<z<1.6$ evolves in size and mass by $\Delta \log \R$ and $\Delta \log \Mstar$ in order to match the local population of galaxies with same velocity dispersion. Clearly high-redshift galaxies must increase both their size and mass. A scenario in which quiescent galaxies do not increase their size over cosmic time is definitively ruled out. The mean growth is $\Delta \log \R = 0.25 \pm 0.05 $ and $\Delta \log \Mstar = 0.16 \pm 0.04 $, corresponding to $\alpha = \Delta \log \R / \Delta \log \Mstar = 1.6 \pm 0.3 $. This can be compared with two examples of growth: $\alpha=1$ (solid orange arrow) for major merging, and $\alpha=2$ (dashed cyan arrow) for minor merging. Although the offset distribution has a shallow peak, compatible with a range in both mass and size growth of $\sim0.5$ dex, the arrow corresponding to $\alpha=1$ is only marginally consistent with the observations. For the sample of $1<z<1.6$ quiescent galaxies, the size growth is steeper than $\alpha=1$ and more consistent with minor merging. In particular, our result is in good agreement with the value $\alpha = 1.60$ found by \citet{nipoti12} for minor merger simulations after averaging over a cosmologically representative set of merger orbits.

\subsection{Evolution at Fixed Ranking in Velocity Dispersion}

We now confront the fact that some of the high-redshift galaxies have velocity dispersions that are larger than any found in the local universe (Figure \ref{fig:sigma_radius_age}). This raises the question of whether our assumption of a constant velocity dispersion is valid, particularly for the population with large \sigmae. Also, numerical simulations do allow a weak evolution in velocity dispersion (see Section \ref{subsec:fixedsigma}). For this reason we refine our matching criterion in the following way. Instead of assuming that the velocity dispersion of an individual galaxy is constant with redshift, we assume that the \emph{ranking} of galaxies in the distribution of  \sigmae\ values is constant. A galaxy at $z>1$ with the largest \sigmae\ will evolve into the galaxy with the largest \sigmae\ in a $z\sim0$ sample drawn from an identical comoving volume. Since the volume probed by our high-redshift survey differs from that probed by the SDSS, we match galaxies at fixed cumulative number density. Such an approach is frequently used to match galaxies at different redshifts \citep[e.g.,][]{behroozi13}, but it is usually applied to the stellar mass function rather than to the velocity dispersion function. Although galaxy mergers can significantly change the stellar mass rank ordering, they will not affect that for the velocity dispersions significantly. We therefore expect our matching procedure to be even more robust.

\citet{bezanson11} have measured the velocity dispersion distribution in both the local universe and up to $z=1.5$. They find that galaxies with very large velocity dispersions are rarer in the local universe and at about $\log \sigmae = 2.40$  they find no evolution in the cumulative distribution, i.e., the number density of galaxies with $\log \sigmae > 2.40$ is constant with redshift (and equals $10^{-4.5} \mathrm{Mpc}^{-3}$). We therefore adopt this velocity dispersion threshold, $\log \sigma_{cr} = 2.40$ (corresponding to 251 km s$^{-1}$), and assume that high-redshift galaxies with $\sigmae > \sigma_{cr}$ will still have $\sigmae > \sigma_{cr}$ at $z\sim0$. Given this large velocity dispersion threshold, incompleteness is not important, since the large-\sigmae\ objects are the ones most easily detected as they are massive and relatively compact. However, some of the high-$\sigmae$ local galaxies are brighter than $r \sim 14.5$, and therefore affected by incompleteness due to saturation and deblending issues \citep{strauss02}. This problem is negligible for the main SDSS sample that we used in the previous analysis, but could be important for the much smaller population of galaxies with $\sigmae > \sigma_{cr}$. To avoid this, we selected a secondary SDSS sample with redshift $0.10 < z < 0.15$ for this analysis. We tested that the results do not change significantly when using the main SDSS sample.

The left panel of Figure \ref{fig:fixed_ranking} shows the mass-size relation for the high and low-redshift samples selected with $\sigmae > \sigma_{cr}$. The high-redshift galaxies are clearly smaller than their local counterparts. The importance of this comparison is that it represents two populations connected by a progenitor-descendent relation: the red points must physically evolve on the mass-size diagram until their distribution is similar to that of the gray points. We infer the growth in size and mass by repeating the procedure described previously: for each high-redshift galaxy we calculate the offset to the local objects, and sum the resulting distribution for the total sample in the right panel of Figure \ref{fig:fixed_ranking}. As before there is unambiguous evolution over $0<z<1.6$. The growth in stellar mass and effective radius for large-\sigmae\ galaxies is more pronounced than that found for the total sample discussed in Section \ref{subsec:fixedsigma}. The mean growth is $\Delta \log \Mstar = 0.34 \pm 0.07$ and $\Delta \log \R = 0.48 \pm 0.08$, with a corresponding $\alpha=1.4\pm0.2$. As before the growth is steeper than $\alpha=1$ (dashed cyan arrow), and minor merging is the preferred growth mechanism.


\section{Summary and Discussion}
\label{sec:discussion}

Using new, deep Keck LRIS spectroscopic data we have measured velocity dispersions for 56 quiescent galaxies at $1<z<1.6$. Taking advantage of public \HST\ imaging and multi-wavelength photometric data, we  derived stellar masses and effective radii. By comparing this sample of high-redshift galaxies with a local sample drawn from the SDSS survey, we find the following results:
\begin{itemize}
\item Quiescent galaxies at high redshift have smaller radii and larger velocity dispersions compared to local objects at fixed stellar mass. However, the offsets in \R\ and \sigmae\ balance each other, and the dynamical masses are similar at low and high redshift, for a given stellar mass.
\item We confirm the applicability at high redshift of the empirical calibration determined at $z\sim0$ by \citet{bezanson11} for deriving inferred velocity dispersions from measured stellar masses and sizes. We find that the velocity dispersions measured with this method have an accuracy of 35\%.
\item We consider a model in which quiescent galaxies evolve over $0<z<1.6$ at fixed velocity dispersion. By using local observations of the velocity dispersion-age relation, we demonstrate that individual galaxies must physically evolve in size and stellar mass in order to match the $z\sim0$ population.
\item In the framework of this model, galaxies evolve at fixed velocity dispersion and increase their effective radii and stellar masses, while their stellar-to-dynamical mass ratio decreases. Quantitatively, we derive a median physical evolution of $\Delta \log \R = 0.25\pm0.05$ and $\Delta \log \Mstar = 0.16\pm0.04$ over $0<z<1.6$ corresponding to a slope in the mass-size plane $\alpha=1.6\pm0.3$. This is consistent with growth via minor merging.
\item For the galaxies with the largest velocity dispersions in our sample, we perform an additional, less restrictive, comparison assuming no evolution in the velocity dispersion ranking. This results in a more convincing and stronger measure of growth, also consistent with minor merging ($\alpha=1.4\pm0.2$).
\item Our spectroscopic data convincingly show that the observed evolution in size and mass over $0<z<1.6$ arises mainly from the physical growth of individual galaxies, and cannot be explained only by progenitor bias.
\end{itemize}

Velocity dispersions represent perhaps one of the most fundamental properties of quiescent galaxies, but accurate measurement at high redshift are observationally challenging. By increasing the initial sample of \citet{newman10} by a factor of 4, in this work we presented the largest sample of velocity dispersion measurements at $z>1$, from which statistically significant conclusions can be drawn. Smaller samples at similar redshifts were obtained by \citet{bezanson13} and \citet{vandesande13}, who also found larger values of \sigmae\ compared to local galaxies of similar stellar mass. By considering evolution in the mass density within a fixed radius, \citet{vandesande13} conclude that quiescent galaxies grow inside-out, in agreement with the minor merging scenario.

By assuming evolution at fixed velocity dispersion, we were able to derive the absolute growth in size and mass for massive quiescent galaxies over $0<z<1.6$. Interestingly, our results agree with the evolution inferred by \citet{vandokkum10} by matching galaxies at fixed number density: for galaxies with $\Mstar \sim 10^{11.2}$ at $z=1.5$ they found $\Delta \log \Mstar = 0.25$ and $\Delta \log \R = 0.5$. The resulting $\alpha = 2$ is consistent with evolution driven by minor merging. It is likewise encouraging that numerical simulations in the framework of $\mathrm{\Lambda}$CDM cosmology succeed in explaining the observed evolution of quiescent galaxies over $0<z<1.5$ in terms of dissipationless merging \citep{nipoti12,cimatti12}. 

An alternative way to study the size growth is to compare the number density of compact objects at low and high redshift. This method does not require velocity dispersions, but relies on number density measurements, for which there are large uncertainties. As a result, different studies have found contradictory results. \citet{trujillo09} and \citet{taylor10nuggets} did not find a population of local old objects as compact as the high-redshift ones, while \citet{poggianti13numberdensity} claim that at least half of the $z > 1$ quiescent galaxies are found at $z \sim 0$ with similar compactness. Moreover, \citet{carollo13} study the evolution of the size function, finding that the size evolution is mainly driven by new arrivals, even though their conclusion is less robust for massive galaxies with $\Mstar > 10^{11}\Msun$. \citet{newman12} consider the minimum physical growth required by the observed evolution of the smallest sizes, and infer a significant physical growth over $ 0 < z < 2 $. They also measure the merger rate of quiescent galaxies, and conclude that for $z < 1$ the rate of minor mergers is large enough to explain the size growth.

Our approach attempts to follow a population of galaxies through cosmic time, thus avoiding the uncertainties involved in the comparison of number densities at high and low redshift. We therefore address the growth of individual galaxies rather than the evolution of the total population. Since the number of massive quiescent galaxies per unit comoving volume increases significantly from $z>1$ to $z=0$ \citep[e.g., by a factor of $\sim3$ according to][]{vanderwel09}, a scenario in which newly quenched galaxies contribute significantly to the size evolution is not inconsistent with our finding of a strong physical growth of the older objects. However, we have demonstrated that progenitor bias cannot be entirely responsible for the size growth.

One further physical process has been proposed for the size growth of quiescent galaxies. A significant mass loss caused by quasar feedback \citep{fan08} or stellar evolution \citep{damjanov09} might in principle induce an adiabatic expansion. In this scenario the velocity dispersion is not conserved, but evolves inversely proportional to the size growth. The comparisons undertaken in this paper cannot test  such a process since, by construction, we assume that velocity dispersions are unchanged during the size growth. However, we note that if the size growth were entirely due to adiabatic expansion, the velocity dispersions at $z\sim1.5$ would be about a factor of two larger than the local ones, at fixed stellar mass \citep{hopkins10}. Our data are in clear disagreement with this prediction (see, e.g., Figure \ref{fig:sigma_radius}), and rule out a dominant role of adiabatic expansion over $0<z<1.6$. 
\\

We acknowledge Carrie Bridge and Kevin Bundy for completing the LRIS observations for two of the slitmasks. The authors recognize and acknowledge the very significant cultural role and reverence that the summit of Mauna Kea has always had within the indigenous Hawaiian community. We are most fortunate to have the opportunity to conduct observations from this mountain.


\appendix

\section{Photometric  Data}
\label{appendix:photometry}

In this appendix we describe in detail the photometric catalogs used to compile the SEDs of the objects in our sample.

\begin{itemize}

\item GOODS-S: we use the catalog from the Multiwavelength Survey by Yale-Chile \citep[MUSYC,][]{cardamone10}, which includes ground-based $U_{38}UBVRIzJHK$, 18 Subaru medium bands in the optical, and the four \spitzer\ IRAC bands.

\item COSMOS: we use data from the NEWFIRM Medium-Band Survey \citep[NMBS,][]{whitaker11}, which consists of deep near-infrared observations in six medium bands taken at the Kitt Peak Mayall 4 m Telescope. We also use Subaru $B_J V_J r^+ i^+ z^+$, Canada-France-Hawaii Telescope (CFHT) $ugrizJHK_S$, 12 Subaru narrow bands, and \spitzer\ IRAC data, all included in the NMBS catalog.

\item EGS: we make use of the catalog released by the WIRCam Deep Survey \citep[WIRDS,][]{bielby12}, consisting in deep CFHT $ugrizJHK_S$ data. To this dataset we add \spitzer\ IRAC data from the Rainbow catalog \citep{barro11}. To avoid inconsistencies due to the difference in aperture correction and zero point between the two catalogs, we determine for each object the ratio of the flux measured by the two surveys in the same band: $f_X = F_{X,{\mathrm WIRDS}} / F_{X,{\mathrm Rainbow}}$, where $X$ is one of the bands that are available in both catalogs ($griJK_S$). We then use $\langle f_X \rangle$, the flux ratio averaged over all the bands,  to correct the Rainbow IRAC fluxes for that object.

\item GOODS-N: three of the five objects of our sample that are located in this field fall in the region covered by the MOIRCS Deep Survey \citep[MODS,][]{kajisawa11}, which includes ground-based $U$, \HST\ $BViz$, Subaru $JHK_S$ and \spitzer\ IRAC data. For the remaining 2 objects we use the data presented in \citet{newman10}: \HST\ $bviz$, Palomar $K_S$ and IRAC.

\item SSA22: for these galaxies we use the Subaru $BVRIZ$ and University of Hawaii 2.2 m telescope $JK$ data described in \citet{newman10}.

\end{itemize}

Additionally, we make use of public data from the \emph{Chandra} and \emph{Spitzer} archives.


\section{Galaxy Structure and Dynamical Masses}
\label{appendix:sersicn}

In Section \ref{sec:dynmasses} we calculated dynamical masses assuming a constant virial factor $\beta=5$. Here we explore the possibility of a varying virial factor and its consequences on the dynamical mass calculation. The virial factor is rigorously constant only if galaxies are assumed to be identical in structure, with just a scaling in mass and size. However, since we used \Sersic\ profiles to describe the surface photometry, we can expect galaxies of different \Sersic\ indices to have different structures and therefore different virial factors. It is possible to derive a theoretical relation between $\beta$ and the \Sersic\ index $n$ for a spherical stellar system with an isotropic velocity dispersion distribution \citep{cappellari06}:
\begin{equation}
\label{eq:beta}
	\beta(n) = 8.87 - 0.831 n + 0.0241 n^2 .
\end{equation}
With this definition of the virial factor, the dynamical mass is
\begin{equation}
\label{eq:mdyn_beta}
	\Mdyn = \frac{\beta(n) \sigmae^2 \R}{G} .
\end{equation}

\begin{figure*}[tbp]
\centering
\includegraphics[width=0.45\textwidth]{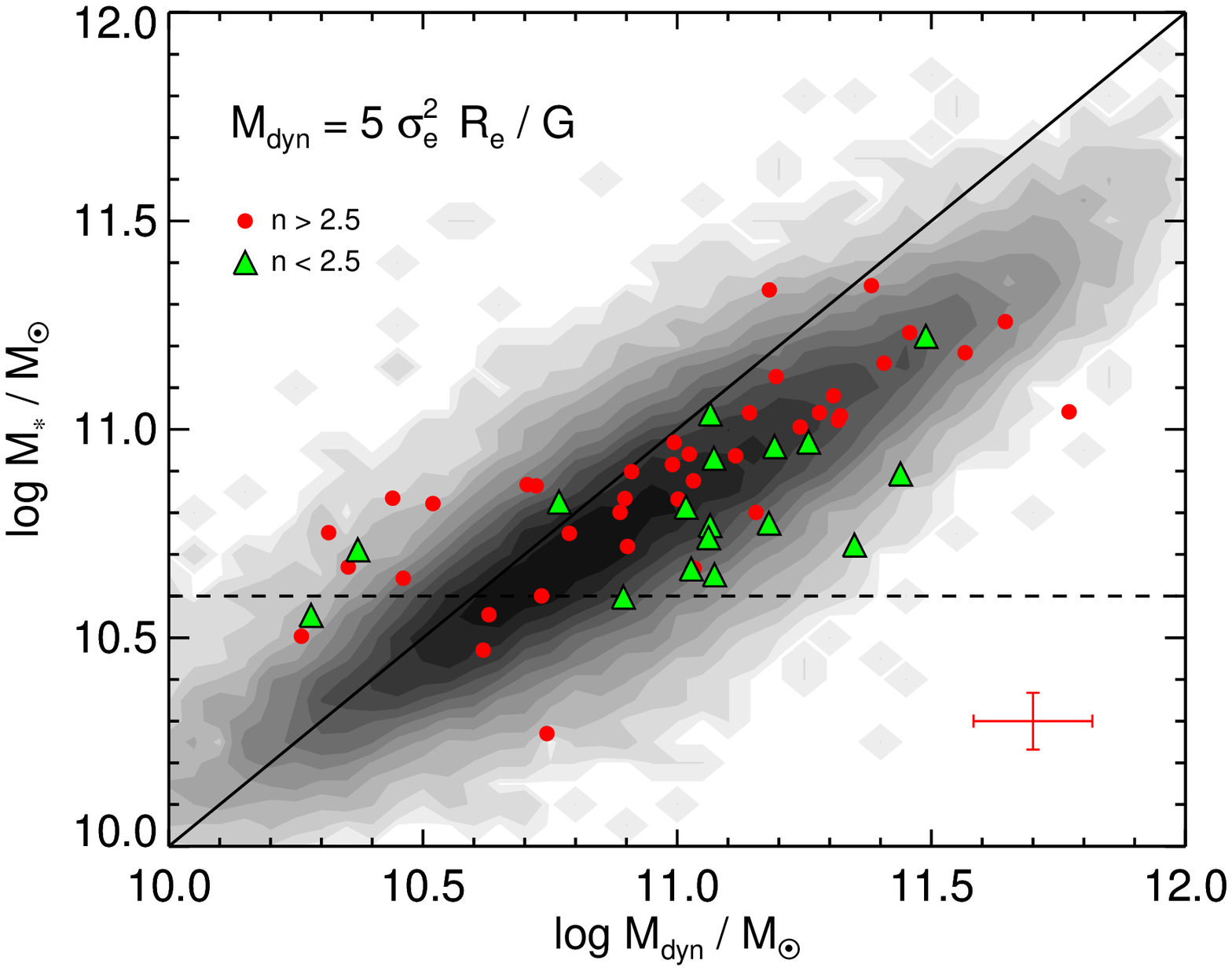}
\includegraphics[width=0.45\textwidth]{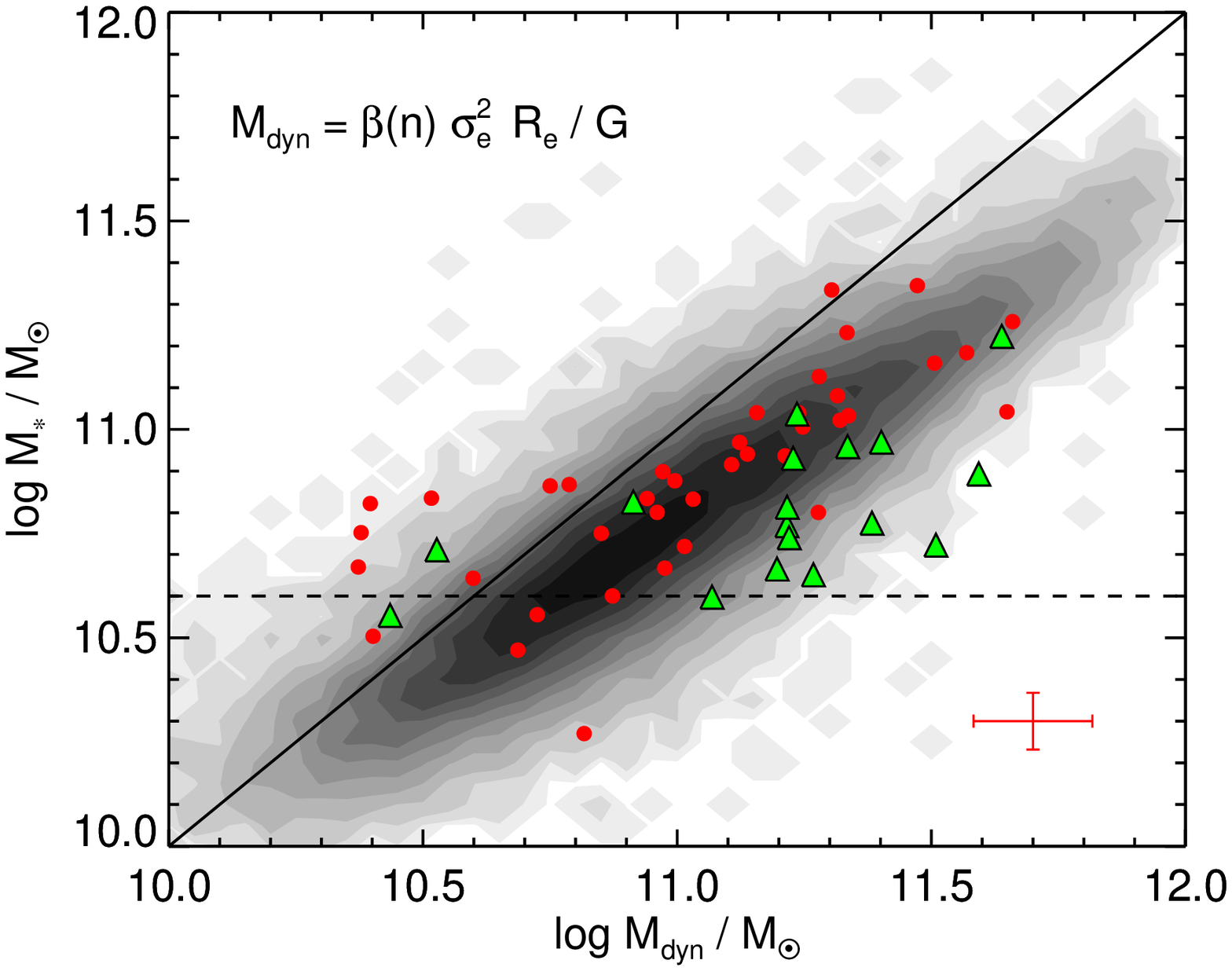}
\caption{Stellar versus dynamical mass for the SDSS sample (grayscale map) and high-redshift galaxies, divided into disks (\Sersic\ index $n<2.5$, green triangles) and spheroidals ($n>2.5$, red points). \emph{Left:} Dynamical masses calculated using a constant virial factor $\beta=5$ (Equation \ref{eq:mdyn}). \emph{Right:} Dynamical masses calculated from Equation \ref{eq:mdyn_beta}, using the \Sersic\ index-dependent virial factor $\beta(n)$.}
\label{fig:mdyn_sersicn}
\end{figure*}

We calculate the virial factor $\beta(n)$ for all the galaxies in our sample, obtaining an average of 6.2 and a standard deviation of 1.1. We then derive dynamical masses according to Equation \ref{eq:mdyn_beta} for the high-redshift galaxies and the SDSS sample, and compare them to the stellar masses in the right panel of Figure \ref{fig:mdyn_sersicn}. In the left panel we show the stellar-dynamical mass comparison using a constant virial factor $\beta=5$, as discussed in Section \ref{sec:dynmasses}. In both panels we plot the objects with $n<2.5$ as green triangles. Although the \Sersic\ index is not a perfect proxy for galaxy structure, low indices are considered a robust indication of the presence of a disk \citep{krajnovic13}. 

From Figure \ref{fig:mdyn_sersicn} we can see a discrepancy between the distribution of high-redshift disk galaxies and the local population. This effect is more pronounced when using a variable virial factor $\beta(n)$. In order to quantify this difference, we bin the SDSS sample in stellar mass, and calculate for each bin the average dynamical mass and its standard deviation. We then compare the distribution of high-redshift disks to the SDSS sequence. If the dynamical masses are calculated with $\beta=5$, then 24\% of the disks (4 out of 17) are outliers, as defined by being more than two standard deviations away from the $z\sim0$ sequence. On the other hand, using $\beta(n)$ yields 47\% outliers (8 out of 17) among disk galaxies. As a comparison, only 18\% of the high-redshift spheroidals (i.e., objects with $n>2.5$) are outliers, independently of which definition of $\beta$ is assumed.

\begin{figure}[bp]
\centering
\includegraphics[width=0.45\textwidth]{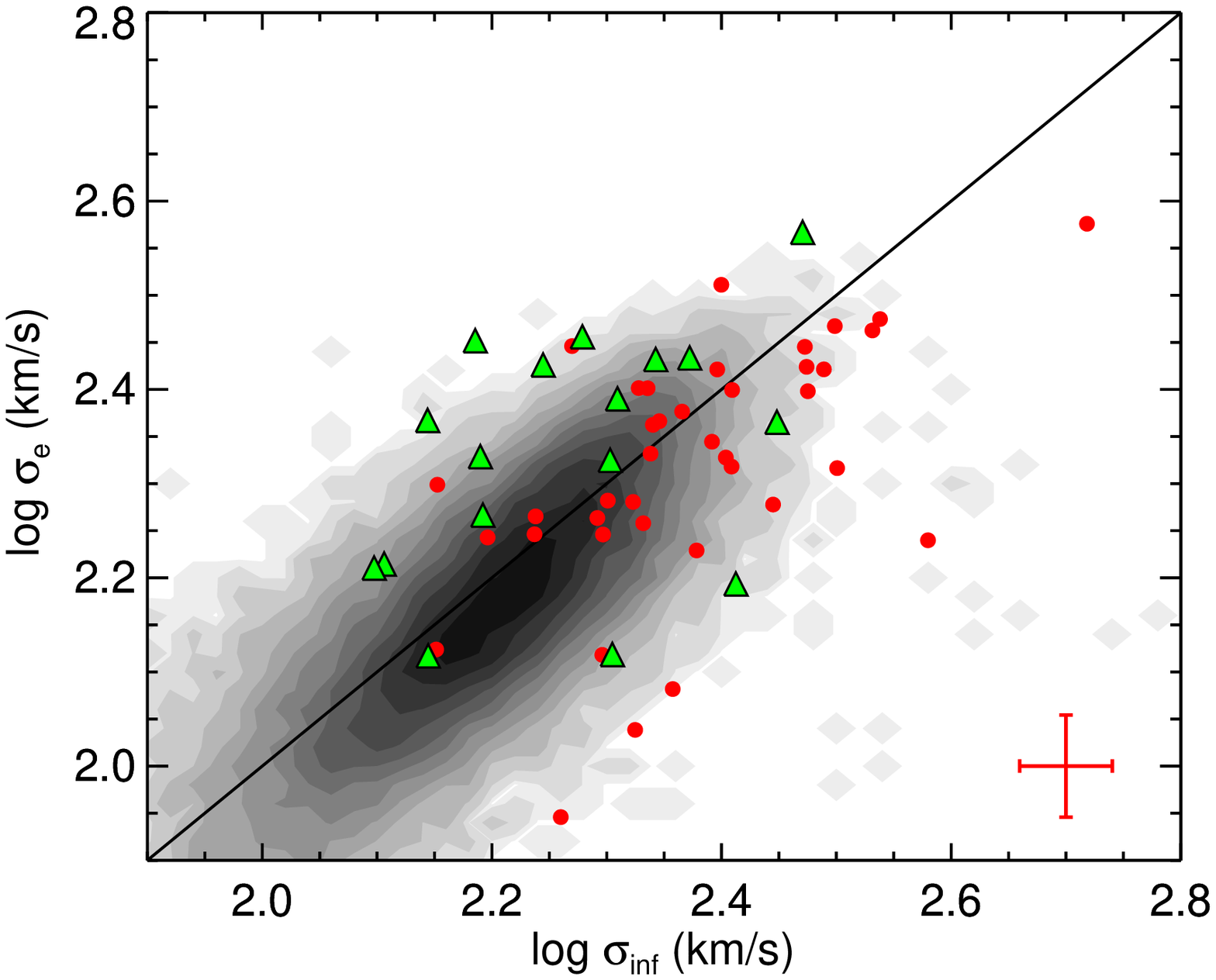}
\caption{Spectroscopically observed versus photometrically inferred velocity dispersions. High-redshift data points are divided into disks ($n<2.5$, green triangles) and spheroidals ($n>2.5$, red points), and the SDSS sample is shown as a grayscale map. All the inferred dispersions are calculated using Equation \ref{eq:sigma_inf_beta}. The median error bars are shown on the bottom right corner.}
\label{fig:sigma_inf_sersic}
\end{figure}

We chose to adopt a constant $\beta=5$ for our analysis because it yields a better agreement between stellar and dynamical masses at both low and high redshift for the full range of \Sersic\ indices. The \Sersic\ index-dependent virial factor $\beta(n)$ seems to be a good description for spheroidal galaxies, but fails to reproduce the stellar-dynamical mass relation for disks. Although this fact is not completely unexpected, since the structure of disks is inherently different from the structure of spheroidals, it is noteworthy that at low redshift there is a good agreement between stellar and dynamical masses, derived using Equation \ref{eq:mdyn_beta}, even for low \Sersic\ indices. This difference might be caused by the fact that the SDSS fibers in most of the cases sample only the central part of a galaxy, measuring the velocity dispersion of the bulge, while at high-redshift we measure the total velocity dispersion, which includes the disk rotation.

Finally, we note that the original definition of inferred velocity dispersion given by \citet{bezanson11} includes the virial factor $\beta(n)$:
\begin{equation}
\label{eq:sigma_inf_beta}
	\sigmainf = \sqrt{ \frac{ G \Mstar }{ 0.557 \beta(n) \R } } .
\end{equation}
We test the agreement between the inferred dispersions derived via this equation and the spectroscopically measured dispersions in Figure \ref{fig:sigma_inf_sersic}. Again, we plot disk galaxies as green triangles. This definition of inferred dispersion produces a good agreement with the \sigmae\ values, with a scatter of 35\%, similar to the one that we obtained using our definition (Equation \ref{eq:sigma_inf}). However, there is a clear trend with the \Sersic\ index, and this method would underpredict the true value of velocity dispersion for most of the $n<2.5$ objects.

\bibliography{dispersions}{}

\end{document}